\documentclass[a4paper,11pt]{article}
\pdfoutput=1 
\usepackage{jheppub} 
\usepackage[T1]{fontenc} 
\usepackage{multirow}
\usepackage[normalem]{ulem}
\usepackage{graphicx}
\usepackage{amsmath,amssymb}
\usepackage{dcolumn}
\usepackage{bm}
\usepackage{color}
\usepackage{subfigure}
\usepackage{appendix}

\newcommand{\f}{\frac}
\newcommand{\lt}{\left}
\newcommand{\m}{m_{\rm P}}
\newcommand{\n}{\nonumber}
\newcommand{\p}{\partial}
\newcommand{\rt}{\right}
\newcommand{\dd}{{\rm d}}

\newcommand{\dt}{\delta}

\newcommand{\ve}{\varepsilon}

\newcommand{\vp}{\varphi}
\newcommand{\pb}{{\rm PBH}}
\newcommand{\cR}{{\cal R}}
\newcommand{\cP}{{\cal P}}

\newcommand{\arxgr}[1]{\href{http://arxiv.org/abs/#1}{{\ttfamily arXiv:#1[gr-qc]}}}
\newcommand{\arxph}[1]{\href{http://arxiv.org/abs/#1}{{\ttfamily arXiv:#1[hep-ph]}}}
\newcommand{\arxth}[1]{\href{http://arxiv.org/abs/#1}{{\ttfamily arXiv:#1[hep-th]}}}

\newcommand{\arxas}[1]{\href{http://arxiv.org/abs/#1}{{\ttfamily arXiv:#1[astro-ph]}}}

\newcommand{\Arxth}[1]{\href{http://arxiv.org/abs/hep-th/#1}{{\ttfamily arXiv:#1[hep-th]}}}
\newcommand{\Arxas}[1]{\href{http://arxiv.org/abs/astro-ph/#1}{{\ttfamily arXiv:#1[astro-ph]}}}

\title{\boldmath An analytical approximation of the evolution of the primordial curvature perturbation in the ultraslow-roll inflation}
\author[a]{Ji-Xiang Zhao}
\author[a]{Nan Li}
\affiliation[a]{Department of Physics, College of Sciences, Northeastern University \\ No. 3-11, Wenhua Road, Shenyang, 110819, China \\}
\emailAdd{2100189@stu.neu.edu.cn}
\emailAdd{linan@mail.neu.edu.cn}

\abstract{Cosmic inflation can enter an ultraslow-roll (USR) stage, if there is a plateau on the inflaton potential. During this stage, the primordial curvature perturbation $\cR_k$ and its power spectrum $\cP_\cR$ can be remarkably enhanced on small scales. In this work, an analytical approximation is provided to systematically study the evolution of $\cR_k$ in the USR inflation. We first discuss the asymptotic solutions of the moduli and arguments of $\cR_k$ and its time derivative $\cR_{k,N}$ on the sub- and super-horizon scales separately and find that all these solutions have simple exponential forms. Then, $\cR_k$ on five typical scales are investigated in order. Our analytical approximation predicts that $\cR_k$ first revolves around the origin in the complex plane, but if it crosses the horizon around the start of the USR stage, there appears a subsequent linear evolution towards or away from the origin. This behavior naturally explains the shape of $\cP_\cR$ from the sharp dip to the peak and matches the numerical results perfectly. Moreover, the minimum of $\cP_\cR$ is exactly proved to be nonvanishing. Our analytical approximation will help the model building in primordial black hole and gravitational wave physics.}

\begin{document}
\maketitle
\flushbottom

\section{Introduction} \label{sec:intro}

The milestone in multi-messenger astronomy is marked by the detection of the gravitational waves from the merger of binary black holes \cite{LIGO}. Some typical characteristics of these black holes, such as their unexpectedly large masses and relatively small spins, are not consistent with usual astrophysical black holes, but are more possible to be of primordial origin \cite{Bird:2016dcv, jp, Clesse}. Primordial black holes (PBHs) are theorized to form before the Big Bang nucleosynthesis, so they are a very powerful probe in the early Universe \cite{Green:2020jor}. For instance, they could explain the strong absorption trough found in the 21-cm global spectrum \cite{Bowman:2018yin, Zhang:2023rnp}, and serve as the seeds of the supermassive black holes in galactic centers \cite{pbh1}, supported by the recent observation from the James Webb Space Telescope on the massive galaxies at high redshifts \cite{2207.09436}. Meanwhile, the first-order scalar perturbations that generate PBHs can also produce the second-order scalar-induced gravitational waves (SIGWs) \cite{Saito:2008jc, Domenech:2021ztg, Ahmed:2021ucx, Kawai:2021edk, Lin:2021vwc, Yi:2020cut, Di:2017ndc}, such as the possible nHz gravitational wave background discovered recently \cite{NANOGrav:2023gor, Antoniadis:2023ott, Reardon:2023gzh, Xu:2023wog, NANOGrav:2023hvm, Inomata:2023zup, Wang:2023ost, Yi:2023mbm, Firouzjahi:2023lzg, Balaji:2023ehk, Franciolini:2023pbf}. More important, PBHs are a natural and promising candidate of dark matter (DM) \cite{dm}.

Unlike astrophysical black holes, the masses of PBHs can range from the Planck mass to supermassive scale ($10^{-38}$--$10^{10}~M_\odot$, with $M_\odot=1.989\times10^{30}~{\rm kg}$ being the solar mass) \cite{Escriva:2022duf}. This versatility enables PBHs to offer valuable insights into different cosmic conundrums \cite{Clesse:2017bsw}. The PBHs with mass $M<5\times 10^{-19}~M_\odot$ have already evaporated because of the Hawking radiation, changing the background intensities of various cosmic rays \cite{zs}. However, those with mass $M>5\times 10^{-19}~M_\odot$ can still exist today, acting as a stable and pressureless candidate of DM. The PBH abundance $f_\pb$ is defined as its proportion in DM today. If $f_\pb\gtrsim 0.1$, the PBHs can be considered as an effective candidate of DM; if $f_\pb\ll 10^{-3}$, its possibility as DM can be safely excluded in the relevant mass range. Currently, according to various constraints on the upper bounds of $f_\pb$ in different mass ranges, there still remains an open mass window from $10^{-17}~M_\odot$ (the asteroid mass range) to $10^{-13}~M_\odot$ (the sub-lunar mass range) that is possible to compose all DM with PBHs \cite{dm, zs, Escriva:2022duf}.

The PBH abundance $f_\pb$ can be calculated from the power spectrum $\cP_\cR(k)$ of the primordial curvature perturbation $\cR$ \cite{PS, peak, Green:2004wb}. On large scales (e.g., the pivot scale $k_* =0.05~{\rm Mpc}^{-1}$ in the Planck satellite experiment), $\cP_\cR$ is precisely measured by the anisotropies of the cosmic microwave background (CMB), with an amplitude of $2.10\times10^{-9}$ \cite{Planck}. However, to produce abundant PBHs, $\cR$ must be large enough, so that $\cP_\cR$ is significantly enhanced up to ${\cal O}(10^{-2})$ on small scales. In the usual single-field slow-roll (SR) inflation models, such a huge enhancement of ${\cal P}_{\cal R}$ is almost impossible. Therefore, the SR conditions must be violated on small scales, and various ultraslow-roll (USR) inflation models are thus proposed \cite{Garcia-Bellido:2017mdw, Kannike:2017bxn, Germani:2017bcs, Motohashi:2017kbs, Dimopoulos:2017ged, Ezquiaga:2017fvi, Ballesteros:2017fsr, Cicoli:2018asa, Ozsoy:2018flq, Dalianis:2018frf, Ballesteros:2018wlw, Cheng:2018qof, Bhaumik:2019tvl, Mahbub:2019uhl, liu, Mishra:2019pzq, Cai:2019bmk, Figueroa:2020jkf, Ragavendra:2020sop, Choi:2021yxz, Kefala:2020xsx, Lyc, Wq, Dalianis:2021iig, Inomata:2021uqj, Cheng:2021lif, Wu:2021mwy, Inomata:2021tpx, Figueroa:2021zah, Hooshangi:2022lao, Geller:2022nkr, Franciolini:2022pav, Balaji:2022zur, Raveendran:2022dtb, Bhatt:2022mmn, Gu:2022pbo, Zjx, Mu:2022dku, Ragavendra:2023ret, Balaji:2022rsy, Qin:2023lgo}. In these models, there is always a plateau or a (near-) inflection point on the inflaton potential, where the inflaton rolls down extremely slowly, amplifying $\cP_\cR$ accordingly.

Albeit the amplitude and shape of $\cP_\cR$ can be straightforwardly obtained in various USR inflation models by numerical methods, the physical comprehension behind seems still indistinct. Previously, in Refs. \cite{Carrilho:2019oqg, liu, Zhai:2023azx}, the spectral index of the steepest growth of ${\cal P}_{\cal R}$ was analytically investigated, and in Ref. \cite{Franciolini:2023lgy}, the authors raised an approximation method to calculate the Fourier mode ${\cal R}_k$ of the primordial curvature perturbation. However, to our knowledge, these analyses are still not sufficient in some aspects. For example, there is always a sharp dip in $\cP_\cR$ before its steep growth. In some early literature \cite{Byrnes:2018txb, Ballesteros:2020sre, Tasinato:2020vdk, Ozsoy:2021qrg, Ozsoy:2021pws, Cole:2022xqc, Karam:2022nym, Pi:2022zxs, shuzhi, Domenech:2023dxx, Alho:2020cdg}, the location of the dip was calculated numerically, but the underlying physical explanation is lacking yet. Although the detail of this dip does not affect the PBH abundance or SIGW spectrum, as they are relevant mainly to the peak of $\cP_\cR$, we still hope to answer whether the minimum of $\cP_\cR$ is zero analytically.

Consequently, this paper is dedicated to study the evolution of $\cR_k$ and the resulting $\cP_\cR$ from a more theoretical perspective, and is a natural extension of our previous numerical results in Ref. \cite{shuzhi}. We will provide an analytical approximation to explore the USR inflation, which will save us away from the mathematical tediousness and extract the physical essence to the most extent. Following our early works in Refs. \cite{Lyc, Wq, Zjx, shuzhi}, by considering an antisymmetric perturbation on the background inflaton potential, inflation can be led into the USR stage smoothly. Our main improvements in this work are threefold. First, by simplifying the two parameters $\ve$ and $\eta$ in the SR and USR stages separately, we are able to obtain the analytical solutions of $\cR_k$ and its time derivative $\cR_{k,N}$. From their asymptotic forms, we achieve $|\cR_k|$, $\theta_{k,N}$, and $\vp_{k,N}$ (with $\theta_{k}$ and $\vp_{k}$ being the arguments of $\cR_k$ and $\cR_{k,N}$) and find that all these quantities possess concise exponential forms, perfectly in agreement with the numerical results in Ref. \cite{shuzhi}. Second, we predict that, besides the revolving evolution of $\cR_k$ around the origin in the complex plane when $k\gg He^N$, there appears an interesting linear evolution of $\cR_k$ towards or away from the origin if $\cR_k$ crosses the horizon around the start of the USR stage, naturally explaining the sharp dip and the peak in $\cP_\cR$. Third, we analytically study the minimum of the dip in $\cP_\cR$ and prove that it cannot reach zero exactly, which is also seldom mentioned before. Altogether, we wish to give a whole picture and thorough understanding of the physical essence in the complicated evolution of $\cR_k$ and the relevant $\cP_\cR$, which will be beneficial to the model building of the USR inflation and help our research on PBH and gravitational wave physics in future.

This paper is organized as follows. The basic equations of motion for $\cR_k$ and $\cR_{k,N}$ are presented in Sec. \ref{sec:PR}. Then, we show our analytical approximation in Sec. \ref{sec:jinsi} and discuss the asymptotic solutions of $|\cR_k|$, $|\cR_k|_{,N}$, $\theta_{k,N}$, and $\vp_{k,N}$ in the sub- and super-horizon limits, respectively. In Sec. \ref{sec:jieguo}, the evolution of $\cR_k$ and the resulting $\cP_\cR$ for five typical scales with different times of horizon-exit are systematically investigated. We conclude in Sec. \ref{sec:con}. We work in the natural system of units and set $c=\hbar=k_{\rm B}=1$.

\section{Basic equations} \label{sec:PR}

In this section, we show the equation of motion for the primordial curvature perturbation ${\cal R}_k$ and its relation to the power spectrum $\cP_\cR(k)$. Two important variables $\theta_k$ and $\vp_k$ are introduced, with their evolutions calculated in detail.

\subsection{USR inflation} \label{sec:basic}

We start from the single-field inflation model, with the corresponding action being
\begin{align}
S=\int\dd^4 x\,\sqrt{-g}\lt[\f{m_{\rm P}^2}{2}R-\frac{1}{2}\p_\mu\phi\p^\mu\phi-V(\phi)\rt], \n
\end{align}
where $m_{\rm P}=1/\sqrt{8\pi G}$ is the reduced Planck mass, $R$ is the Ricci scalar, $\phi$ is the inflaton field, and $V(\phi)$ is its potential. The inflaton potential $V$ can be further decomposed into its background $V_{\rm b}(\phi)$ and a perturbation $\dt V(\phi)$ on it. In this work, we choose $V_{\rm b}$ as the Kachru--Kallosh--Linde--Trivedi potential \cite{KKLT},
\begin{align}
V_{\rm b}(\phi)=V_0\f{\phi^2}{\phi^2+(\m/2)^2}. \n
\end{align}
Meanwhile, we consider a perturbation $\dt V$ on $V_{\rm b}$,
\begin{align}
\dt V(\phi)=-A\f{\phi -\phi _0}{1+ ( \phi -\phi _0)^2/ ( 2\sigma^2)}, \n
\end{align}
where three model parameters $A$, $\phi_0$, and $\sigma$ characterize the amplitude, position, and width of $\dt V$, respectively. (In principle, the specific form of $\dt V$ is not unique, and other kinds of $\dt V$ can be found in Refs. \cite{Lyc, Wq, Zjx}.) Hence, $\dt V$ is antisymmetric around $\phi_0$, so it can be smoothly connected on $V_{\rm b}$ on both sides of $\phi_0$. In addition, we demand $A=V_{{\rm b},\phi}(\phi_0)$ for simplicity, in order to create a perfect plateau around $\phi_0$, leading inflation into the USR stage. Following Ref. \cite{Wq}, we choose the model parameters as $V_0/\m^4=10^{-10}$, $\phi_0/\m=1.31$, and $\sigma/\m=0.0831881$, such that there can be PBHs with mass $M\approx 10^{-17}~M_\odot$ and abundance $f_{\rm PBH}\approx 0.1$. Moreover, the initial conditions for inflation are set to be $\phi/\m=3.30$ and $\phi_{,N}/\m=-0.0137$, so as to satisfy the Planck constraints on the power spectrum and the tensor-to-scalar ratio on the CMB pivot scale $k_*$ \cite{Planck}.


In cosmic inflation, it is more convenient to utilize the number of $e$-folds $N$ as the time variable, defined as $\dd N=H\,\dd t=\dd\ln a$, where $t$ is the cosmic time, $a(t)=e^N$ is the scale factor, and $H(t)=\dot{a}/a$ is the Hubble expansion rate. Furthermore, to characterize the motion of the inflaton, two useful parameters are introduced as
\begin{align}
\ve&=-\f{\dot H}{H^2}=\f{\phi_{,N}^2}{2\m^2}, \label{ve}\\
\eta&=-\f{\ddot\phi}{H\dot\phi}=\frac{\phi_{,N}^2}{2m_{\rm P}^2}-\frac{\phi_{,NN}}{\phi_{,N}}. \label{eta}
\end{align}
In terms of these parameters, the Friedmann equation for cosmic expansion becomes
\begin{align}
H^2=\frac{V}{(3-\ve)m^2_{\rm P}}, \label{Fried}
\end{align}
and the Klein--Gordon equation for the inflaton field can be reexpressed as
\begin{align}
\phi_{,NN}+(3-\ve)\phi_{,N}+\frac{1}{H^2}V_{,\phi}=0. \label{kgphi}
\end{align}

\begin{figure}[htb]
\centering \includegraphics[width=0.65\linewidth]{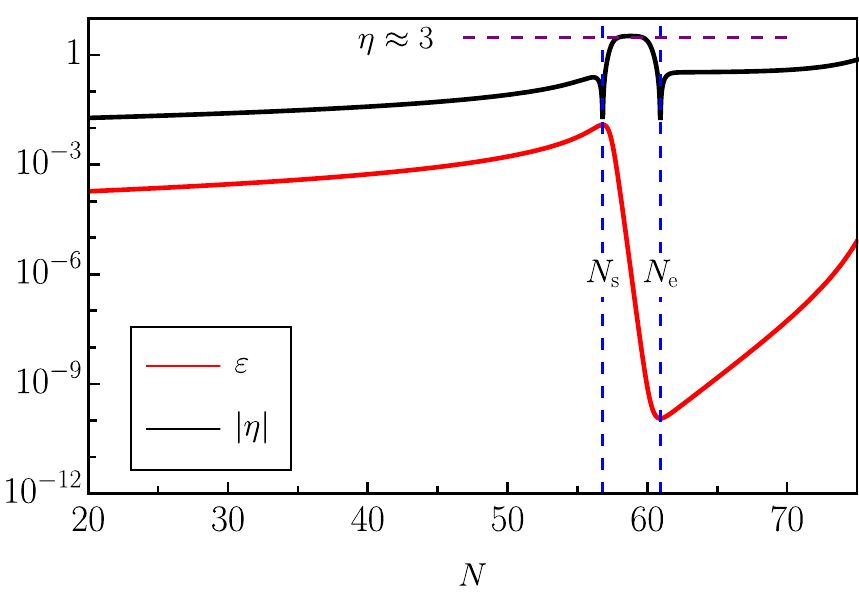}
\caption{The evolutions of the parameters $\ve$ and $|\eta|$ as a function of the number of $e$-folds $N$. The starting and ending points of the USR stage are determined by $\eta(N_{\rm{s}})=\eta(N_{\rm{e}})=0$, with $N_{\rm s}=56.81$ and $N_{\rm e}=60.93$. During the USR stage, $\ve$ decreases nearly exponentially, but $|\eta|$ increases significantly and maintains a value around 3 for a sufficiently long period.} \label{canshu}
\end{figure}

In the usual SR inflation, both $\ve$ and $|\eta|$ are always much smaller than 1 and are thus named as the SR parameters. However, in the USR stage, $\ve$ drops almost exponentially and $|\eta|$ may even approach ${\cal O}(1)$, so the SR conditions are broken. Thus, the starting and ending points of the USR stage can be determined by $\eta(N_{\rm{s}})=\eta(N_{\rm{e}})=0$. With the model parameters listed above, we have $N_{\rm s}=56.81$ and $N_{\rm e}=60.93$. The evolutions of $\ve$ and $|\eta|$ in both SR and USR stages are shown in Fig. \ref{canshu}. Moreover, the number of $e$-folds when the relevant scale crosses the horizon is denoted as $N_{\rm out}$ (i.e., when $k=He^{N_{\rm out}}$), which is an important quantity for the subsequent discussions in Sec. \ref{sec:jieguo}.

\subsection{${\cal R}_k$ and ${\cal P}_{\cal R}$}

Now, we move on to the perturbations on the background spacetime. In Newtonian gauge, the perturbed metric reads
\begin{align}
\dd s^2&=-(1+2\Psi)\,\dd t^2+a^2(1-2\Psi)\dt_{ij}\,\dd x^i\dd x^j, \n
\end{align}
where $\Psi$ is the scalar perturbation. (Here, we have neglected the vector perturbation, tensor perturbation, and anisotropic stress, as they are irrelevant to PBH formation.) A more convenient and gauge-invariant scalar perturbation is the primordial curvature perturbation $\cR$ defined as
\begin{align}
{\cal R}=\Psi+\frac{H}{\dot{\phi}}\,\dt\phi=\Psi+\frac{\dt\phi}{\phi_{,N}}, \n
\end{align}
and the equation of motion for its Fourier mode ${\cal R}_k$ is the Mukhanov--Sasaki (MS) equation \cite{Mukhanov, Sasaki},
\begin{align}
\cR_{k,NN}+(3+\ve-2\eta)\cR_{k,N}+\lt(\frac{k}{He^{N}}\rt)^2\cR_k=0. \label{MS}
\end{align}

The initial conditions for the MS equation can be fixed by the Bunch--Davies vacuum \cite{BD}. In the very early Universe, both $\epsilon$ and $|\eta|$ are very small, and the asymptotic solution of $\cR_k$ is
\begin{align}
\cR_k=\f{e^{ik/(He^N)}}{z\sqrt{2k}}\lt(1+\f{iHe^N}{k}\rt), \label{vk}
\end{align}
where $z=\phi_{,N}e^N$ (i.e., $z^2=2m_{\rm P}^2a^2\ve$). At the initial time, we have $k\gg He^N$, so from Eq. (\ref{vk}), we obtain
\begin{align}
\cR_k\bigg|_{N\to N_{\rm{ini}}}&=\f{e^{ik/(He^N)}}{z\sqrt{2k}}, \label{crki}\\
\cR_{k,N}\bigg|_{N\to N_{\rm{ini}}}&=-\f{e^{ik/(He^N)}}{z\sqrt{2k}}\lt(\f{z_{,N}}{z}+\f{ik}{He^N}\rt), \label{crkni}
\end{align}
where $N_{\rm{ini}}$ is the initial number of $e$-folds.

Furthermore, the PBH abundance and the SIGW spectrum can be calculated from the power spectrum of $\cR$, which corresponds to the two-point correlation function of $\cR_k$ in the Fourier space. The statistical information and observational significance of $\cR_k$ is encoded in the dimensionless power spectrum ${\cal P}_{\cal R}(k)$ as
\begin{align}
\lt.{\cal P}_{\cal R}(k)=\f{k^3}{2\pi^2}|{\cal R}_{k}|^2\rt|_{k\ll aH}. \n
\end{align}
We should point out that, in the usual SR inflation, $\cR_k$ becomes almost frozen when the relevant scale crosses the horizon, and ${\cal P}_{\cal R}$ can be effectively evaluated at the epoch of horizon-exit. Nonetheless, in the USR inflation, $\cR_k$ can still evolve significantly after the horizon-exit, so ${\cal P}_{\cal R}$ must be determined at the end of inflation.

Generally speaking, both $\cR_k$ and ${\cal P}_{\cal R}$ can be obtained by numerically solving Eqs. (\ref{ve})--(\ref{MS}), but this is always a time-consuming task. Consequently, one of the aims of this work is to explore an appropriate and analytical approximation to calculate $\cR_k$ and ${\cal P}_{\cal R}$. Only in this way can we extract the physical essence in the complicated evolution of $\cR_k$ and the characteristic shape of ${\cal P}_{\cal R}$ in the USR inflation, without getting lost in the tedious calculation program.

\subsection{$\theta_k$ and $\varphi_k$} \label{sec:jiaodu}

As the MS equation is complex, it is more convenient to decompose $\cR_k$ as
\begin{align}
\cR_k=|\cR_k| e^{-i\theta_k}, \label{cRt}
\end{align}
where $|\cR_k|$ and $\theta_k$ are the modulus and argument of $\cR_k$. Another advantage of this decomposition, rather than decomposing $\cR_k$ into the real and imaginary parts, is that we will observe interesting revolving and linear evolutions of $\cR_k$ in the complex plane, to be shown in Sec. \ref{sec:jieguo}.

Substituting Eq. (\ref{cRt}) into Eq. (\ref{MS}), we arrive at the equations of motion for $|\cR_k|$ and $\theta_k$,
\begin{align}
|\cR_k|_{,NN}+(3+\ve-2\eta)|\cR_k|_{,N}+\lt(\f{k}{He^{N}}\rt)^2\lt[1-\lt(\f{He^{N}\theta_{k,N}}{k}\rt)^2\rt]|{\cal R}_k|=0, \label{shibu}\\
\theta_{k,NN}+\lt(3+\ve-2\eta+2\f{|\cR_k|_{,N}}{|\cR_k|}\rt)\theta_{k,N}=0. \label{xubu}
\end{align}
Furthermore, from Eq. (\ref{crki}), the initial conditions for $|\cR_k|$ read
\begin{align}
|\cR_k|\bigg|_{N\to N_{\rm{ini}}}&=-\f{1}{z\sqrt{2k}}, \label{rki}\\
|\cR_k|_{,N}\bigg|_{N\to N_{\rm{ini}}}&=\f{z_{,N}}{z^2\sqrt{2k}}. \label{rkin}
\end{align}
When the inflaton $\phi$ rolls down from its potential $V$, we have $z=\phi_{,N}e^N<0$. Since $|\cR_k|$ is positive, we have assigned a negative sign in Eq. (\ref{rki}).

Below, for our analytical purpose, we provide two mathematical tricks to simplify the calculations of $|\cR_k|$ and $\theta_k$. First, from Eqs. (\ref{vk}) and (\ref{cRt}), we can obtain the evolution of the argument $\theta_k$ as \cite{Kefala:2020xsx}
\begin{align}
\theta_{k,N}=\f{1}{2z^2 He^N |\cR_k|^2}, \label{th}
\end{align}
and the detailed derivation can be found in \ref{appendix}. Hence, $\theta_{k,N}$ is positive-definite, so $\theta_k$ is a monotonic function.

Second, from Eq. (\ref{cRt}), we have $\cR_{k,N}=|\cR_k|_{,N}e^{-i\theta_k}-i|\cR_k|\theta_{k,N}e^{-i\theta_k}$, so the evolution of the modulus $|\cR_k|$ is
\begin{align}
|\cR_k|_{,N}=\sqrt{|\cR_{k,N}|^2-(|\cR_k|\theta_{k,N})^2}. \label{bianhuan}
\end{align}
By this means, we are able to achieve $|\cR_k|$ and $\theta_k$ more quickly, as both Eqs. (\ref{th}) and (\ref{bianhuan}) are first-order differential equations, in which $\cR_{k}$ and $\cR_{k,N}$ can be easily obtained from the MS equation via the analytical approximation in Sec. \ref{sec:jinsi}. In contrast, it is impossible to solve Eq. (\ref{shibu}) analytically, due to the complicity in the prefactor of $|\cR_k|$.

Similarly, we decompose the derivative of $\cR_k$ as
\begin{align}
\cR_{k,N}=|\cR_{k,N}| e^{-i\vp_k}, \n
\end{align}
where the argument $\vp_k$ can be obtained from Eq. (\ref{cRt}) as
\begin{align}
\vp_k =\theta_k+\arctan\lt(\frac{|\cR_k|}{|\cR_k|_{,N}}\theta_{k,N}\rt). \label{vp}
\end{align}
Substituting Eq. (\ref{th}) into Eq. (\ref{vp}), we arrive at the evolution of $\vp_k$ as
\begin{align}
\vp_{k,N}=\f{1}{2z^2He^N|\cR_k|^2}-\f{(2z^2He^N|\cR_k||\cR_k|_{,N})_{,N}}{(2z^2He^N|\cR_k||\cR_k|_{,N})^2+1}. \label{vpn}
\end{align}
Different from $\theta_{k,N}$ in Eq. (\ref{th}), $\vp_{k,N}$ depends not only on $|\cR_k|$, but also on $|\cR_k|_{,N}$.

In summary, Eqs. (\ref{cRt}), (\ref{th}), (\ref{bianhuan}), and (\ref{vpn}) summarize the evolutions of the moduli and arguments of $\cR_k$ and $\cR_{k,N}$, which are the starting point of our following discussions.

\section{Analytical approximation} \label{sec:jinsi}

In this section, we propose our analytical approximation to calculate $|\cR_k|$ and $|\cR_k|_{,N}$ in two different cases: $k\gg He^N$ and $k\ll He^N$, in which the scale of $\cR_k$ is sub- and super-horizon, respectively. Next, we bring the results into Eqs. (\ref{th}) and (\ref{vpn}) to obtain $\theta_{k,N}$ and $\vp_{k,N}$. Our main results are summarized in Tabs. \ref{table:1} and \ref{table:2}.

We start our analytical approximation from the two parameters $\ve$ and $\eta$. First, from Fig. \ref{canshu}, in the SR stage when $N<N_{\rm s}$, $\ve$ is nearly invariant and $\eta$ is very small, so we approximately have
\begin{align}
\ve\approx\ve(N_{\rm s})=\ve_{\rm s}, \quad \eta\approx0. \n
\end{align}
Second, in the USR stage when $N_{\rm s}<N<N_{\rm e}$, as the inflaton potential is extremely flat, from Eqs. (\ref{eta}) and (\ref{kgphi}), we arrive at $\eta\approx3$. Then, from Eqs. (\ref{ve}) and (\ref{eta}), $\eta$ can be reexpressed as $\eta=\ve-{\ve_{,N}}/(2\ve)$, so we approximately obtain
\begin{align}
\ve\approx \ve_{\rm s} e^{-6(N-N_{\rm s})}, \quad \eta\approx3. \n
\end{align}
Third, when $N>N_{\rm e}$, the inflation dynamics significantly depends on the specific inflaton potential, so it is hard to have a general approximation method any more, and we will not consider it in this work.

Before exploring Eq. (\ref{MS}), we still need to make two assumptions. First, $\eta$ and $H$ are assumed to be constant. Second, $\ve$ is neglected when it is added to $\eta$. These assumptions are valid in both SR and USR stages, as can be evidently seen from Fig. \ref{canshu}. With these preparations, Eq. (\ref{MS}) can be exactly solved as a function of $N$ as
\begin{align}
\cR_k&= e^{-\lt(\f{3}{2}-\eta\rt)N}\lt[A J_{\f{3}{2}-\eta}\lt(\f{k}{He^N}\rt)+B J_{-\f{3}{2}+\eta}\lt(\f{k}{He^N}\rt)\rt], \label{jie}\\
\cR_{k,N}&=e^{-\lt(\f{5}{2}-\eta\rt)N}\f kH \lt[-A J_{\f{1}{2}-\eta}\lt(\f{k}{He^N}\rt) +B J_{-\f{1}{2}+\eta}\lt(\f{k}{He^N}\rt)\rt], \label{daoshu}
\end{align}
where $J_\alpha(x)$ is the Bessel function of the first kind, and the coefficients $A$ and $B$ are the functions of $k$, $H$, and $\eta$, but are independent of $N$. Also, we should mention that both $A$ and $B$ are complex numbers, since ${\cal R}_k$ itself is complex.

\subsection{$k\gg He^N$} \label{sec:dayu}

We first consider the sub-horizon case with $k\gg He^N$. Under this circumstance, our analytical approximation is relatively simple, because $J_{\alpha}(x)\sim1/\sqrt{x}$ when $x=k/(He^N)\gg1$, irrespective of the value of $\alpha$. As a result, from Eqs. (\ref{jie}) and (\ref{daoshu}), we easily obtain
\begin{align}
|\cR_k|&\approx c_1 e^{-(1-\eta)N}\sim
\begin{cases}
e^{-N} & (\mbox{when $\eta\approx0$}), \\
e^{2N} & (\mbox{when $\eta\approx3$}),
\end{cases} \label{r}\\
|\cR_{k,N}|&\approx c_2 e^{-(2-\eta)N}\sim
\begin{cases}
e^{-2N} & (\mbox{when $\eta\approx0$}), \\
e^{N} & (\mbox{when $\eta\approx3$}),
\end{cases} \label{rn}
\end{align}
where the coefficients $c_1$ and $c_2$ can be obtained from $A$ and $B$, but the explicit expressions are unnecessary here. We observe that the time-dependence of $\cR_k$ and $\cR_{k,N}$ is rather simple. More interpretation of these results will be presented in Sec. \ref{sec:jieguo}.

Now, we discuss the SR and USR stages separately. In the SR case, $\eta\approx0$, so from Eqs. (\ref{r}) and (\ref{rn}), we have
\begin{align}
|\cR_k|\sim e^{-N}, \quad |\cR_{k,N}|\sim e^{-2N}. \label{zhongjian}
\end{align}
Substituting $|\cR_k|$ into Eq. (\ref{th}) and considering $\ve\approx\ve_{\rm s}$, we obtain the evolution of $\theta_k$ as
\begin{align}
\theta_{k,N}\sim\f{1}{e^{2N}\ve_{\rm s} He^N e^{-2N}}\sim e^{-N}. \label{xiaoyu0theta}
\end{align}
Then, substituting Eqs. (\ref{zhongjian}) and (\ref{xiaoyu0theta}) into Eq. (\ref{bianhuan}), we have
\begin{align}
|\cR_k|_{,N}\sim e^{-2N}. \n
\end{align}
For the evolution of $\vp_k$, substituting Eqs. (\ref{zhongjian}) and (\ref{xiaoyu0theta}) into Eq. (\ref{vpn}), we obtain
\begin{align}
\vp_{k, N}\sim\frac{1}{e^{2N}\ve_{\rm s} He^N e^{-2N}}\sim e^{-N}. \n
\end{align}

Similarly, in the USR case, $\eta\approx3$, so
\begin{align}
|\cR_k|\sim e^{2N}, \quad |\cR_{k,N}|\sim e^N. \n
\end{align}
Taking into account $\ve\approx \ve_{\rm s} e^{-6(N-N_{\rm s})}$, we have
\begin{align}
\theta_{k,N}\sim\f{1}{e^{2N}\ve_{\rm s}e^{-6(N-N_{\rm s})} He^N e^{4N}}\sim e^{-N}. \label{xiaoyu3theta}
\end{align}
Thus, $\theta_{k,N}$ has the same time-dependence as that in Eq. (\ref{xiaoyu0theta}). However, this happens only for the sub-horizon case. In the super-horizon case, $\theta_{k,N}$ may have totally different results in the SR and USR cases, to be shown in Sec. \ref{sec:xiaoyu}. Furthermore, by the same means, we obtain
\begin{align}
|\cR_k|_{,N}\sim e^{N}, \n
\end{align}
and
\begin{align}
\vp_{k, N}\sim\frac{1}{e^{2N}\ve_{\rm s}e^{-6(N-N_{\rm s})} He^N e^{4N}}\sim e^{-N}. \n
\end{align}
It is interesting to find that $\vp_{k,N}$ behaves the same in both SR and USR stages, but this is understandable. In the numerator of the second term in Eq. (\ref{vpn}), $2z^2He^N|\cR_k||\cR_k|_{,N}$ is actually constant no matter $\eta\approx 0$ or 3. Thus, its derivative with respect to $N$ vanishes, so $\vp_{k,N}\approx\theta_{k,N}\sim e^{-N}$.

All the above results are summarized in Tab. \ref{table:1}, indicating that, when a scale is sub-horizon, $|\cR_k|$ is a monotonic function of $N$. Therefore, $\cR_k$ either revolves towards the origin in the complex plane at a decelerated rate in the SR stage, or away from the origin at an accelerated rate in the USR stage. These conclusions will be compared with the numerical results in Sec. \ref{sec:jieguo}.\footnote{Here, two subtle points should be mentioned. First, the results of $|\cR_k|_{,N}$ seem inconsistent with those of $|\cR_k|$, but we must point out that $|\cR_k|_{,N}$ cannot be trivially obtained from the derivative of $|\cR_k|$, but must be calculated from Eq. (\ref{bianhuan}). Second, one cannot naively expect $|\cR_k|_{,N}\sim |\cR_{k,N}|$ (although their dependence on $N$ does coincide), but must use Eq. (\ref{bianhuan}) in principle. These two points also apply in Sec. \ref{sec:xiaoyu}.}

\begin{table}[htb]
\renewcommand\arraystretch{1.25}
\centering
\begin{tabular}{m{1.2cm}<{\centering}|m{1.8cm}<{\centering}|m{1.6cm}<{\centering}|m{1.6cm}<{\centering}|m{1.6cm}<{\centering}}
\hline\hline
$\eta$ & $|\cR_k|$ & $\theta_{k,N}$ & $|{\cal R}_k|_{,N}$ & $\vp_{k,N}$ \\
\hline
0 & $e^{-N}$ & $e^{-N}$ & $e^{-2N}$ & $e^{-N}$ \\
\hline
3 & $e^{2N}$ & $e^{-N}$ & $e^{N}$ & $e^{-N}$ \\
\hline\hline
\end{tabular}
\caption{All the possible results of $|\cR_k|$, $\theta_{k,N}$, $|\cR_k|_{,N}$, and $\vp_{k,N}$ in the sub-horizon case with $k\gg He^N$ via our analytical approximation. The parameter $\eta\approx 0$ or 3 corresponds to the SR or USR stage, respectively. From the behaviors of $|\cR_k|$ and $\theta_{k,N}$, $\cR_k$ either revolves towards the origin at a decelerated rate in the SR stage, or away from the origin at an accelerated rate in the USR stage.} \label{table:1}
\end{table}

\subsection{$k\ll He^N$} \label{sec:xiaoyu}

Next, we turn to the super-horizon case with $k\ll He^N$. The current situation is a little bit complicated than the sub-horizon case, and the essential reason is that, when $x=k/(He^N)\ll 1$, the asymptotic form of the Bessel function $J_\alpha(x)\sim x^\alpha$ depends on $\alpha$. Consequently, from Eqs. (\ref{jie}) and (\ref{daoshu}), we obtain
\begin{align}
|\cR_k|&\approx\big|c_3 e^{-(3-2\eta)N}+c_4\big|=\begin{cases}
\big|c_3e^{-3N}+c_4\big| & \mbox{(when $\eta\approx0$)}, \\
\big|c_3e^{3N}+c_4\big| & \mbox{(when $\eta\approx3$)},
\end{cases} \label{dayu}\\
|\cR_{k,N}|&\approx \big|c_5e^{-(3-2\eta)N}+c_{6}e^{-2N}\big|=\begin{cases}
\big|c_5e^{-3N}+c_{6}e^{-2N}\big| & \mbox{(when $\eta\approx0$)}, \\
\big|c_5e^{3N}+c_{6}e^{-2N}\big| & \mbox{(when $\eta\approx3$)},
\end{cases} \label{dayun}
\end{align}
where the coefficients $c_3$ to $c_6$ can also be obtained from $A$ and $B$. Now, the situation becomes more intractable, as there are two competitive terms in both $|\cR_k|$ and $|\cR_k|_{,N}$. Hence, we must be very careful to determine the leading-order one, and we discuss this issue in detail below.

We start from the SR stage with $\ve\approx\ve_{\rm s}$ and $\eta\approx0$. If the leading-order terms are
\begin{align}
|\cR_k|\approx c_3 e^{-3N}, \quad |\cR_{k,N}|\approx c_5e^{-3N}, \n
\end{align}
from Eqs. (\ref{th}), (\ref{bianhuan}), and (\ref{vpn}), by the same method in Sec. \ref{sec:dayu}, we obtain
\begin{align}
\theta_{k,N}\sim e^{3N}, \quad |\cR_k|_{,N}\sim e^{-3N}, \quad \vp_{k, N}\sim e^{3N}. \n
\end{align}
Similarly, if the leading-order terms are
\begin{align}
|\cR_k|\approx c_4, \quad |\cR_{k,N}|\approx c_6 e^{-2N}, \n
\end{align}
the corresponding results read
\begin{align}
\theta_{k,N}\sim e^{-3N}, \quad |\cR_k|_{,N}\sim e^{-2N}, \quad \vp_{k, N}\sim e^{-N}. \n
\end{align}

Next, we move on to the USR stage with $\ve\approx\ve_{\rm s} e^{-6(N-N_{\rm s})}$ and $\eta\approx3$. If the leading-order terms are
\begin{align}
|\cR_k|\approx c_3 e^{3N}, \quad |\cR_{k,N}|\approx c_5 e^{3N}, \n
\end{align}
we obtain
\begin{align}
\theta_{k,N}\sim e^{-3N}, \quad |\cR_k|_{,N}\sim e^{3N}, \quad \vp_{k, N}\sim e^{-3N}. \n
\end{align}
Similarly, if the leading-order terms are
\begin{align}
|\cR_k|\approx c_4, \quad |\cR_{k,N}|\approx c_6 e^{-2N}, \n
\end{align}
we have
\begin{align}
\theta_{k,N}\sim e^{3N}, \quad |\cR_k|_{,N}\sim e^{-2N}, \quad \vp_{k,N}\sim e^{-5N}. \n
\end{align}

Again, all the above results are summarized in Tab. \ref{table:2}, according to the increasing rate of $|\cR_k|$. Because of the competition between the two terms in Eqs. (\ref{dayu}) and (\ref{dayun}), there are altogether four different combinations at present. From the last two rows, we find that, during the USR stage, $\vp_{k,N}$ decreases dramatically, so $\vp_k$ will reach a constant very soon. Under these circumstances, $\cR_k$ will have tiny angular velocity in the complex plane, so there will appear a linear evolution of $\cR_k$, to be shown in Figs. \ref{fig:5289}--\ref{fig:5884}.

\begin{table}[htb]
\renewcommand\arraystretch{1.25}
\centering
\begin{tabular}{m{1.5cm}<{\centering}|m{1.6cm}<{\centering}m{1.6cm}<{\centering}|m{1.6cm}<{\centering}m{1.6cm}<{\centering}}
\hline\hline
$\eta$ & $|\cR_k|$ & $\theta_{k,N}$ & $|\cR_k|_{,N}$ & $\vp_{k,N}$\\
\hline
0 & $e^{-3N}$ & $e^{3N}$ & $e^{-3N}$ & $e^{3N}$ \\
0 & const.     & $e^{-3N}$ & $e^{-2N}$ & $e^{-N}$ \\
\hline
3 & const.     & $e^{3N}$ & $e^{-2N}$  & $e^{-5N}$ \\
3 & $e^{3N}$  & $e^{-3N}$  & $e^{3N}$ & $e^{-3N}$ \\
\hline\hline
\end{tabular}
\caption{Same as Tab. \ref{table:1}, but in the super-horizon case with $k\ll He^N$. The current situation is more complicated, due to the competition between the two terms in Eqs. (\ref{dayu}) and (\ref{dayun}), so all the possible four combinations of the leading-order terms are considered.}
\label{table:2}
\end{table}

Last, we briefly comment that our analytical approximation mainly applies to the two limiting cases with $k\gg He^N$ and $k\ll He^N$, but is unfortunately not suitable for the intermediate region around $N\sim N_{\rm out}$ with $k\sim He^N$, because at this time, it is difficult to determine the value of $[k/(He^N)]^2$ in front of $\cR_k$ in Eq. (\ref{MS}).

\section{Evolution of the primordial curvature perturbation} \label{sec:jieguo}

\begin{figure}[htb]
\centering \includegraphics[width=0.65\linewidth]{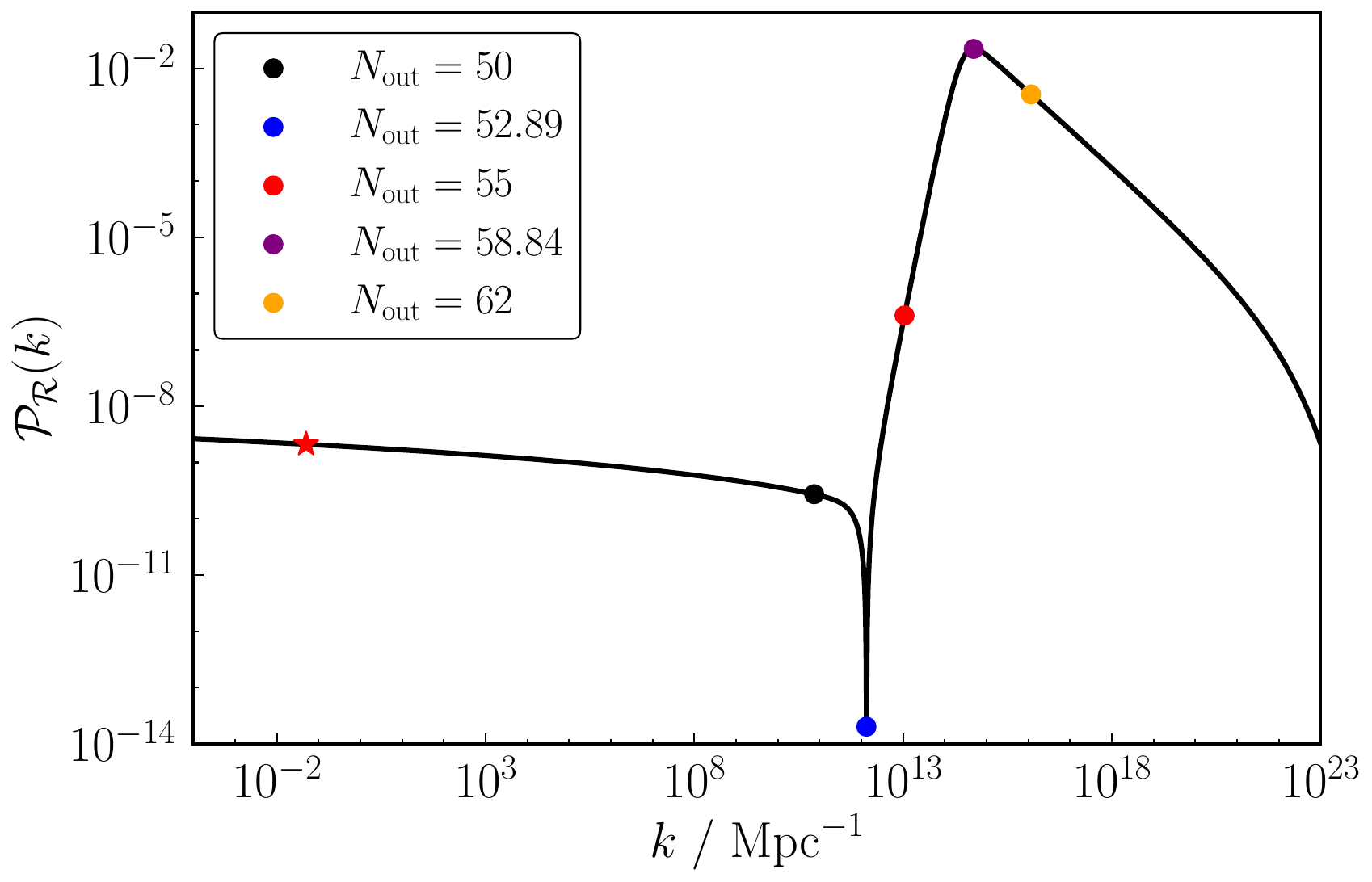}
\caption{The power spectrum $\cP_{\cR}(k)$ in the USR inflation, with the model parameters in Sec. \ref{sec:basic}. On large scales (e.g., the CMB pivot scale $k_*=0.05~{\rm Mpc}^{-1}$ denoted with an asterisk), $\cP_{\cR}$ is nearly scale-invariant with an amplitude of $2.10\times 10^{-9}$ \cite{Planck}. On small scales, $\cP_{\cR}$ can be enhanced up to ${\cal O}(10^{-2})$ to produce abundant PBHs and SIGWs. Five typical scales with different $N_{\rm out}=50$, 52.89, 55, 58.84, and 62 are marked as dots with different colors, corresponding to the nearly scale-invariant region, sharp dip, steep
growth, peak, and falling stage, respectively. The minimum of $\cP_\cR$ in the sharp dip is merely $1.61\times 10^{-14}$, but not zero.} \label{PR}
\end{figure}

In this section, we utilize the results in Sec. \ref{sec:jinsi} to provide a whole picture of the evolution of the primordial curvature perturbation $\cR_k$ in the USR inflation. Meanwhile, we compare our analytical approximation with the numerical results. For such comparison, five typical scales are chosen, for which the numbers of $e$-folds when they cross the horizon are $N_{\rm out}=50$, 52.89, 55, 58.84, and 62, respectively.
The reason for choosing these numbers is that they correspond to five typical positions in the power spectrum $\cP_\cR(k)$: the nearly scale-invariant region, sharp dip, steep growth, peak, and falling stage, respectively, as shown in Fig. \ref{PR}.

\begin{figure*}[htb]
\begin{center}
\subfigure[]{\includegraphics[width=0.48\linewidth]{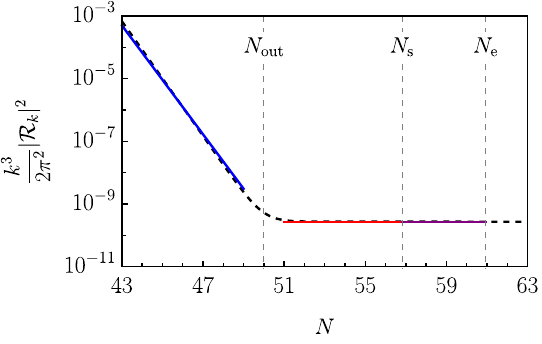} \label{501}}
\subfigure[]{\includegraphics[width=0.46\linewidth]{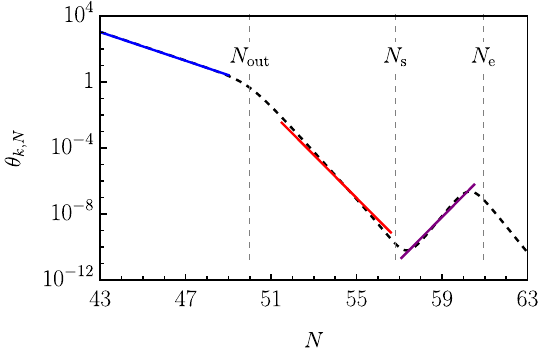} \label{502}} \\
\subfigure[]{\includegraphics[width=0.48\linewidth]{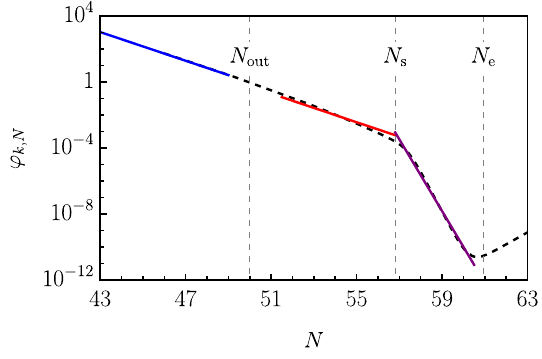} \label{503}}
\subfigure[]{\includegraphics[width=0.48\linewidth]{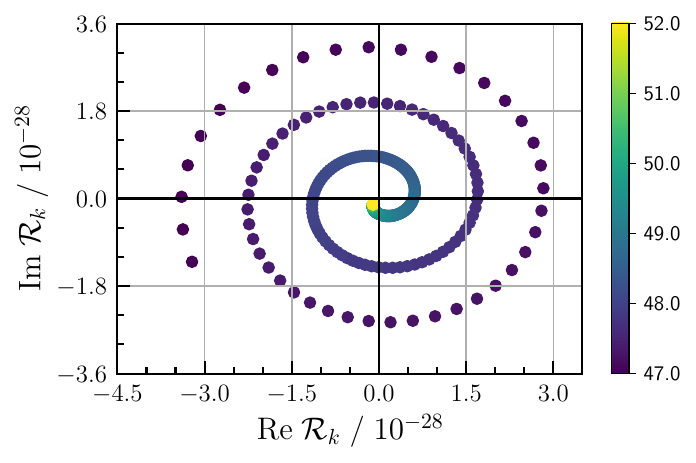} \label{504}}
\end{center}
\caption{The evolutions of the primordial curvature perturbation $\cR_k$ and the two arguments $\theta_k$ and $\vp_k$ with $N_{\rm out}=50$. Before $N_{\rm out}$, we have $k\gg He^N$ and $\eta\approx0$, and $k^3|\cR_k|^2/(2\pi^2)\sim e^{-2N}$, $\theta_{k,N}\sim e^{-N}$, and $\vp_{k,N}\sim e^{-N}$, so the slopes of the blue lines in Figs. \ref{501}--\ref{503} are $-2$, $-1$, and $-1$. Similarly, when $N_{\rm out}<N<N_{\rm s}$, we have $k\ll He^N$ and $\eta\approx0$, so the slopes of the red lines are 0, $-3$, and $-1$. When $N_{\rm s}<N<N_{\rm e}$, we have $k\ll He^N$ and $\eta\approx 3$, so the slopes of the purple lines are 0, 3, and $-5$. All these values match the numerical results (the dashed lines) very well. Moreover, the evolution of $\cR_k$ from $N=47$ to 52 ($\Delta N=0.01$) is shown in the complex plane in Fig. \ref{504}, in which $\cR_k$ first revolves around the origin at a decelerated rate and finally stops near it. This means that, for the large-scale curvature perturbation that crosses the horizon much earlier before the USR stage, $|\cR_k|$ becomes almost constant once $N>N_{\rm out}$, and the influence from the USR inflation is negligible.} \label{fig:50}
\end{figure*}

\begin{figure*}[htb]
\begin{center}
\subfigure[]{\includegraphics[width=0.48\linewidth]{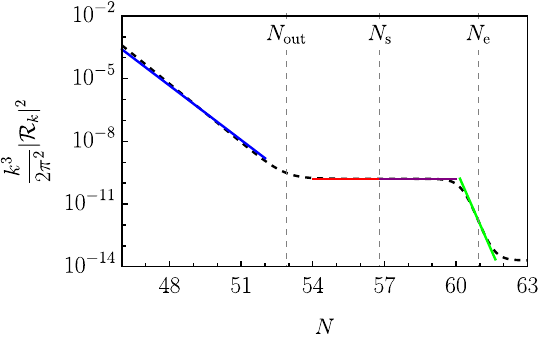}\label{52891}}
\subfigure[]{\includegraphics[width=0.46\linewidth]{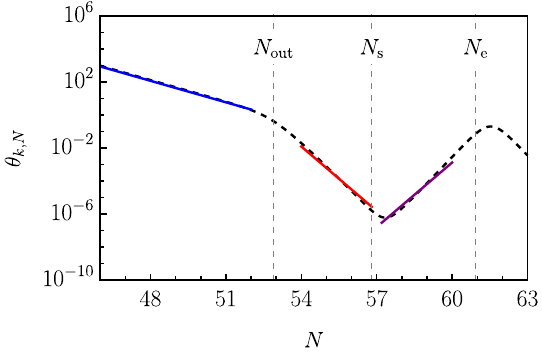}\label{52892}} \\
\subfigure[]{\includegraphics[width=0.48\linewidth]{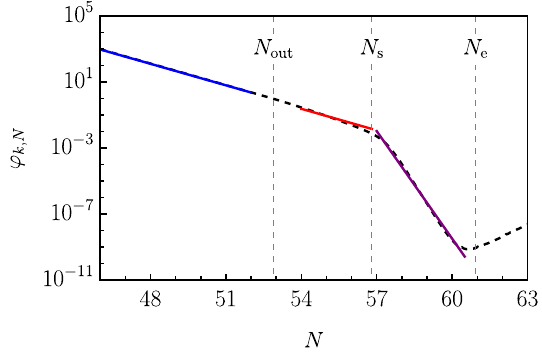}\label{52893}}
\subfigure[]{\includegraphics[width=0.48\linewidth]{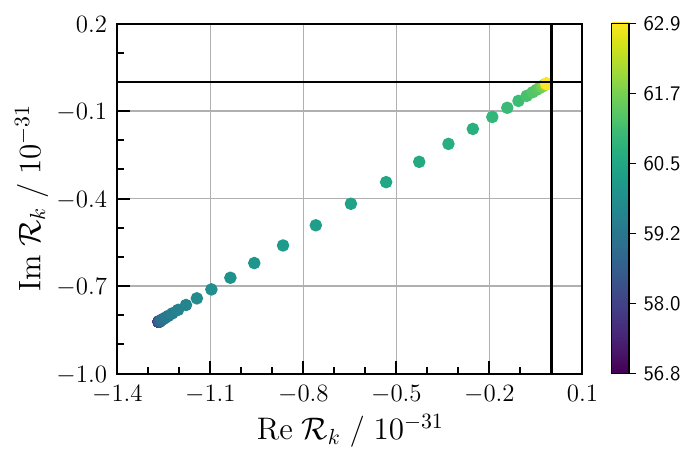}\label{52894}}
\end{center}
\caption{Same as Fig. \ref{fig:50}, but with $N_{\rm out}=52.89$. Before the end of the USR stage, the evolutions of $|\cR_k|$, $\theta_{k}$, and $\vp_{k}$ basically resemble those in Fig. \ref{fig:50}. However, around $N_{\rm e}$, ${\cal R}_k$ is no longer frozen but decreases dramatically as $|{\cal R}_k|\sim e^{-3N}$, and $\vp_{k,N}$ is extremely small at the same time, so there appears a new linear evolution of $\cR_k$ after the revolving process (not shown in the figure) towards but not through origin, as shown in Fig. \ref{52894} with $N=56.8$ to 62.9 ($\Delta N=0.1$). This linear evolution naturally explains the sharp dip in ${\cal P}_{\cal R}$, and the minimum of the dip is not exactly zero from both theoretical and numerical points of view.} \label{fig:5289}
\end{figure*}

\begin{figure*}[htb]
\begin{center}
\subfigure[]{\includegraphics[width=0.48\linewidth]{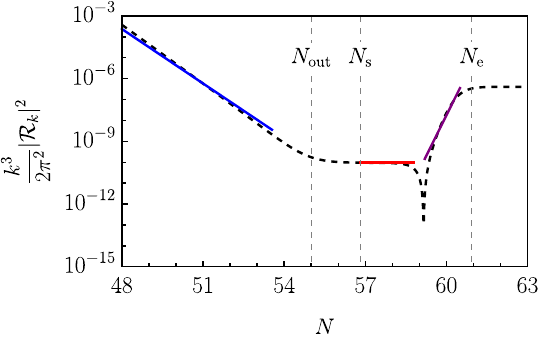}\label{551}}
\subfigure[]{\includegraphics[width=0.46\linewidth]{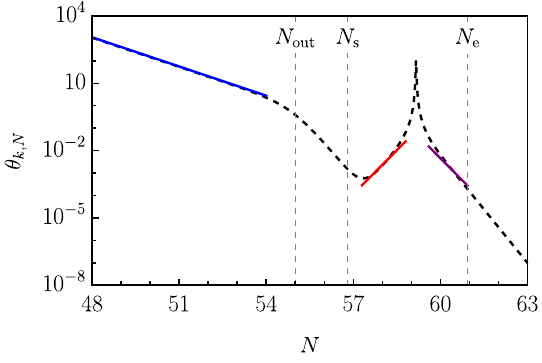}\label{552}} \\
\subfigure[]{\includegraphics[width=0.48\linewidth]{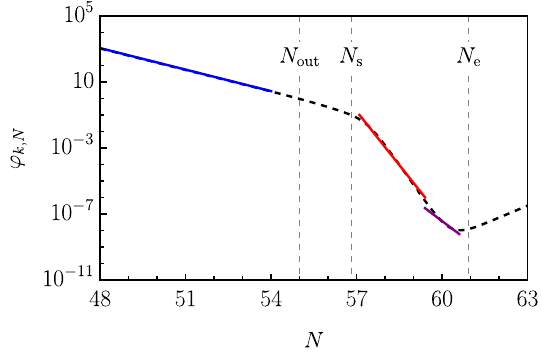}\label{553}} \subfigure[]{\includegraphics[width=0.48\linewidth]{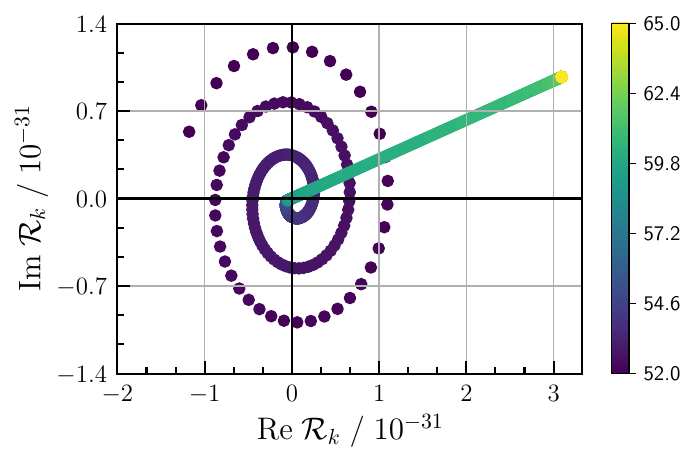}\label{554}}
\end{center}
\caption{
Same as Fig. \ref{fig:50}, but with $N_{\rm out}=55$. When $N< N_{\rm out}$, the evolutions of $|\cR_k|$, $\theta_{k}$, and $\vp_{k}$ are analogous to those in Fig. \ref{fig:50}, as shown in the blue lines in Figs. \ref{551}--\ref{553}. During the USR stage, we have $k\ll He^N$ and $\eta\approx3$, but there are two competitive terms in $|\cR_k|$ from Eq. (\ref{dayu}). As a result, in the early phase, we have $|\cR_k|\sim{\rm const.}$, $\theta_{k,N}\sim e^{3N}$, and $\vp_{k,N}\sim e^{-5N}$, so the slopes of the red lines are 0, $3$, and $-5$. Similarly, in the late phase, the slopes of the purple lines are 6, $-3$, and $-3$. Hence, ${\cal R}_k$ shows a linear evolution away from the origin and finally stops at a distant point in the complex plane, explaining the steep growth in $\cP_\cR$, as shown in Fig. \ref{554} with $N=52$ to 65 ($\Delta N= 0.01$). Moreover, at $N\approx 59.15$, a sharp decrease appears in $|{\cal R}_k|$ in Fig. \ref{551}, accompanied by a spike in $\theta_{k,N}$ in Fig. \ref{552}, as a natural consequence from Eq. (\ref{th}).} \label{fig:55}
\end{figure*}

\begin{figure*}[htb]
\centering
\subfigure[]{\includegraphics[width=0.48\linewidth]{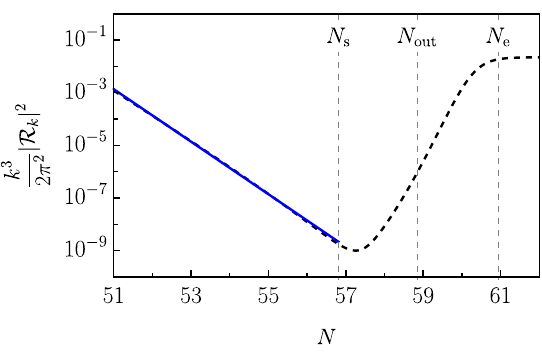}\label{58841}}
\subfigure[]{\includegraphics[width=0.46\linewidth]{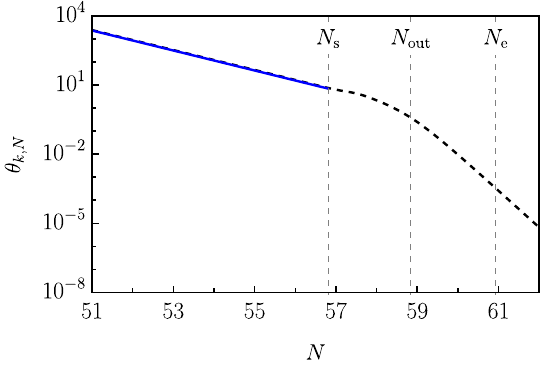}\label{58842}} \\
\subfigure[]{\includegraphics[width=0.48\linewidth]{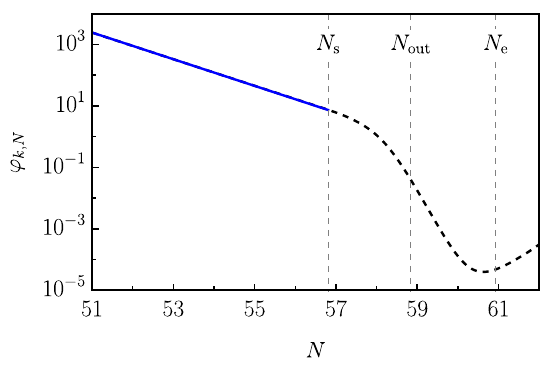}\label{58843}}
\subfigure[]{\includegraphics[width=0.5\linewidth]{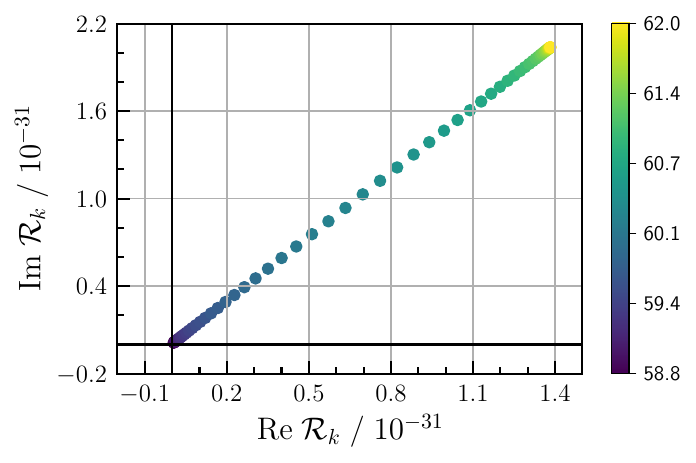}\label{58844}}
\caption{Same as Fig. \ref{fig:50}, but with $N_{\rm out}=58.84$. Here, our analytical approximation applies only to the situation when $N<N_{\rm s}$ with $k\gg He^N$. As shown in the blue lines in Figs. \ref{58841}--\ref{58843}, the evolutions of $|\cR_k|$, $\theta_k$, and $\vp_k$ are similar to the previous cases. From the dashed lines in Figs. \ref{58841} and \ref{58843}, the rapid increase of $|\cR_k|$ and decrease of $\vp_{k,N}$ almost maintain as before in Figs. \ref{551} and \ref{553}, so the linear evolution of $\cR_k$ still exists. In Fig. \ref{58844}, $\cR_k$ linearly moves away from the origin and eventually stops at a rather distant point, resulting in the peak in $\cP_\cR$ (the previous revolving process is not shown), with $N=58.8$ to $62$ ($\Delta N=0.05$).} \label{fig:5884}
\end{figure*}

\begin{figure*}[htb]
\centering
\subfigure[]{\includegraphics[width=0.48\linewidth]{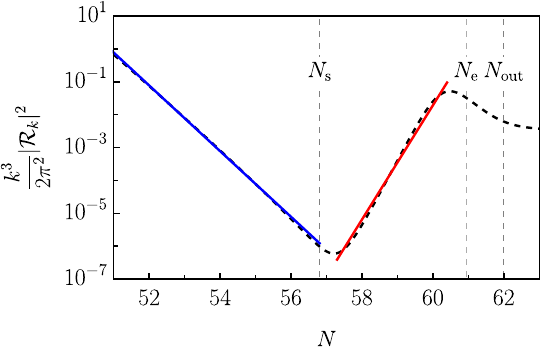}\label{621}}
\subfigure[]{\includegraphics[width=0.46\linewidth]{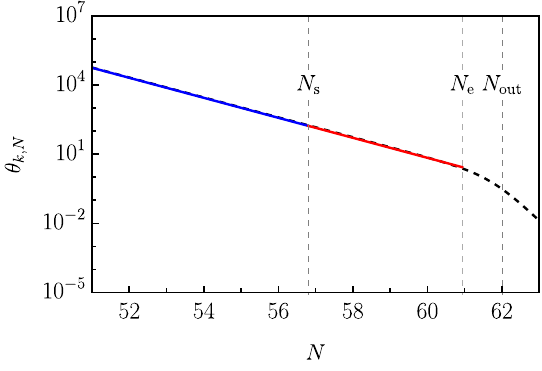}\label{622}} \\
\subfigure[]{\includegraphics[width=0.48\linewidth]{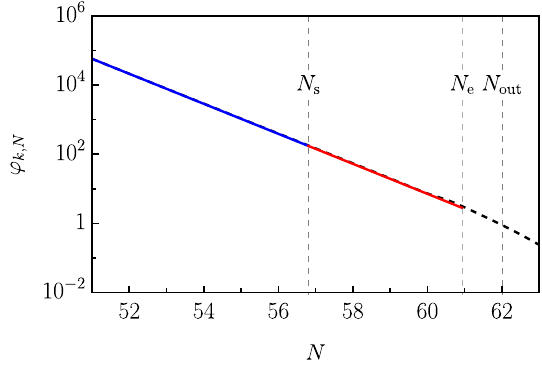}\label{623}}
\subfigure[]{\includegraphics[width=0.5\linewidth]{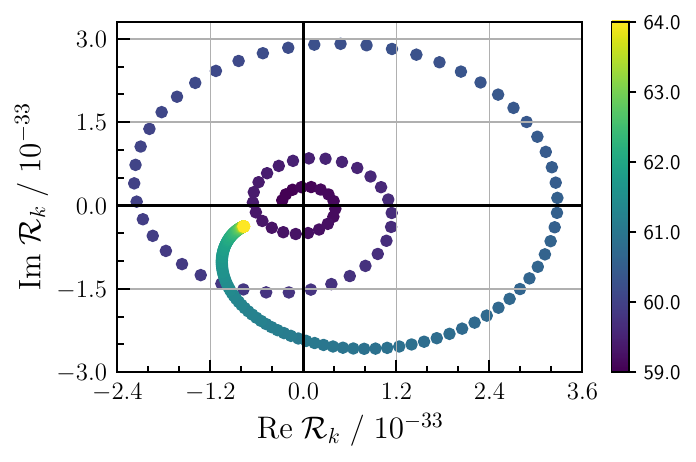}\label{624}}
\caption{Same as Fig. \ref{fig:50}, but with $N_{\rm out}=62$. Because $N_{\rm out}>N_{\rm e}$, we only need to consider the simple case with $k\gg He^N$. When $N<N_{\rm s}$, we have $\eta\approx0$, so the slopes of the blue lines in Figs. \ref{621}--\ref{623} are $-2$, $-1$, and $-1$. Next, when $N_{\rm s}<N<N_{\rm e}$, we have $\eta\approx3$, so the slopes of the red lines are $4$, $-1$, and $-1$. Moreover, ${\cal R}_k$ first revolves towards the origin and then away from it, as can be seen in Fig. \ref{624} with $N=59$ to 64 ($\Delta N=0.02$). Although $|{\cal R}_k|$ changes dramatically around the USR stage, the value of $\vp_{k,N}$ is still large enough, so ${\cal R}_k$ cannot show a linear evolution in the complex plane any more.} \label{fig:62}
\end{figure*}

Below, the evolution of ${\cal R}_k$ with different $N_{\rm out}$ is systematically explored in order and compared with the numerical results (i.e., the lower right panels in Figs. \ref{fig:50}--\ref{fig:62}) in our previous work in Ref. \cite{shuzhi}.

First, we start from the case with $N_{\rm out}=50$, in which the relevant scale crosses the horizon much earlier than $N_{\rm s}$, and our results are shown in Fig. \ref{fig:50}. Before $N_{\rm out}$, we have $k\gg He^N$ and $\eta\approx 0$. Therefore, $|\cR_k|\sim e^{-N}$ [or equivalently, $k^3|\cR_k|^2/(2\pi^2)\sim e^{-2N}$], $\theta_{k,N}\sim e^{-N}$, and $\vp_{k,N}\sim e^{-N}$ (the first row in Tab. \ref{table:1}), as depicted in the blue lines in Figs. \ref{501}--\ref{503}. However, after $N_{\rm out}$, we should be cautious and distinguish the SR and USR stages. When $N_{\rm out}<N<N_{\rm s}$ (i.e., the SR stage), we have $k\ll He^N$ and $\eta\approx 0$, so $|\cR_k|\sim{\rm const.}$, $\theta_{k,N}\sim e^{-3N}$, and $\vp_{k,N}\sim e^{-N}$ (the second row in Tab. \ref{table:2}). However, when $N_{\rm s}<N<N_{\rm e}$ (i.e., the USR stage), we have $k\ll He^N$ and $\eta\approx 3$, so $|\cR_k|\sim{\rm const.}$, $\theta_{k,N}\sim e^{3N}$, and $\vp_{k,N}\sim e^{-5N}$ (the third row in Tab. \ref{table:2}). These behaviors can also be found in the red and purple lines in Figs. \ref{501}--\ref{503}. Altogether, we observe that our analytical approximation fits the numerical results (the dashed lines) perfectly in all these situations. Furthermore, the evolution of $\cR_k$ from $N=47$ to 52 is exhibited in the complex plane in Fig. \ref{504}, with the difference of the number of $e$-folds between two adjacent dots set to be $\Delta N=0.01$. From the results of $|\cR_k|$ and $\theta_{k,N}$, it is easy to find that $\cR_k$ revolves clockwise towards the origin and eventually stops near it, where $|\cR_k|$ tends to be constant. The interval between the dots monotonically decreases, meaning that $\cR_{k}$ decelerates gradually. All these behaviors indicate that the USR stage has negligible effect on such large-scale curvature perturbation, and $|\cR_k|$ becomes almost frozen once $N>N_{\rm out}$, which is reflected in the nearly scale-invariant region in $\cP_\cR$.

Second, we study the scale with $N_{\rm out}=52.89$, and our results are shown in Fig. \ref{fig:5289}. In fact, most behaviors of $|\cR_k|$, $\theta_{k,N}$, and $\vp_{k,N}$ [e.g., the slopes of the blue, red, and purple lines in Figs. \ref{52891}--\ref{52893}] remain the same as those in the previous case. The unique distinction is that, from the first row in Tab. \ref{table:2}, around the end of the USR stage, $|\cR_k|\sim e^{-3N}$ decreases dramatically [the green line in Fig. \ref{52891}], and $\vp_{k,N}$ is extremely small. Hence, there appears a new linear evolution of ${\cal R}_k$ towards but not through the origin subsequent to the revolving process, as shown in Fig. \ref{52894}, with $N = 56.8$ to 62.9. This interesting point naturally explains the sharp dip in ${\cal P}_{\cal R}$. Moreover, we should emphasize that this dip actually cannot reach zero, consistent with the numerical result [the dashed line in Fig. \ref{52891}]. If there were $\cR_k$ that eventually stops at the origin, we would have $|\cR_k|=0$ and $\theta_{k,N}=0$ simultaneously. However, from Eq. (\ref{th}), this is strictly forbidden, so the minimum of $\cP_\cR$ can never be zero, contrary to the naive inference in Ref. \cite{Goswami:2010qu}.

Third, we consider the scale with $N_{\rm out}=55$, which is before but near $N_{\rm s}$. When $N<N_{\rm out}$, the situation is analogous to the former two cases, as shown in the blue lines in Figs. \ref{551}--\ref{553}. However, when $N_{\rm s}<N<N_{\rm e}$, we have $k\ll He^N$ and $\eta\approx 3$, so we have to face the two competitive terms in Eq. (\ref{dayu}) and determine the leading-order one in advance. As a result, in the early phase of the USR stage, we follow the third row in Tab. \ref{table:2}, so $|\cR_k|\sim{\rm const.}$, $\theta_{k,N}\sim e^{3N}$, and $\vp_{k,N}\sim e^{-5N}$, corresponding to the red lines. In contrast, in the late phase of the USR stage, we arrive at the last row in Tab. \ref{table:2}, so $|\cR_k|\sim e^{3N}$, $\theta_{k,N}\sim e^{-3N}$, and $\vp_{k,N}\sim e^{-3N}$, as shown with the purple lines. As $|{\cal R}_k|$ is significantly amplified and $\vp_{k,N}$ is tiny simultaneously, ${\cal R}_k$ shows a linear evolution away from the origin and finally terminates at a remote point in the complex plane, inducing the steep growth in ${\cal P}_{\cal R}$, as shown Fig. \ref{554} with $N=52$ to 65. In addition, the competition between the two terms in Eq. (\ref{dayu}) causes a sharp decrease of $|{\cal R}_k|$ at $N\approx59.15$. From Eq. (\ref{th}), this decrease also explains a spike in $\theta_{k,N}$, as can be seen in Figs. \ref{551} and \ref{552} (such divergent behaviors are beyond our analytical approximation).

Fourth, the peak of the power spectrum $\cP_\cR$ corresponds to the scale with $N_{\rm out}=58.84$, and our analytical approximation is partly suitable to this special case. Now, $N_{\rm out}$ is almost in the middle between $N_{\rm s}=56.81$ and $N_{\rm e}=60.93$, but the approximation is inapplicable when $k\sim He^N$. Therefore, we can only find the similar behaviors of $|\cR_k|$, $\theta_{k,N}$, and $\vp_{k,N}$ to those in the previous three cases when $N<N_{\rm s}$ (not $N_{\rm out}$, since it is larger than $N_{\rm s}$ now). During the USR stage, the numerical method is still needed. From the dashed lines in Figs. \ref{58841} and \ref{58843}, the rapid increase of $|\cR_k|$ and decrease of $\vp_{k,N}$ almost maintain as before in Figs. \ref{551} and \ref{553}, so the linear evolution of $\cR_k$ still exists, resulting in the peak in $\cP_\cR$ eventually, as shown in Fig. \ref{58844} with $N=58.8$ to $62$.

Last, we finish our discussion with $N_{\rm out}=62$. Because $N_{\rm out}>N_{\rm e}$ at present, we only need to take into account the simple case with $k\gg He^N$. Therefore, in the SR stage, we have $\eta\approx 0$, so $|{\cal R}_k|\sim e^{-N}$, $\theta_{k,N}\sim e^{-N}$, and $\vp_{k,N}\sim e^{-N}$ (the first row in Table \ref{table:1}), as shown in the blue lines in Figs. \ref{621}--\ref{623}. Next, during the USR stage, we have $\eta\approx 3$, so $|{\cal R}_k|\sim e^{2N}$, $\theta_{k,N}\sim e^{-N}$, and $\vp_{k,N}\sim e^{-N}$ (the second row in Tab. \ref{table:1}), as shown in the red lines. Since $|\cR_k|$ first decreases and then increases in the current situation, ${\cal R}_k$ first revolves towards the origin and then away from it, as can be seen in Fig. \ref{624} with $N=59$ to 64. Besides, there is no more linear evolution of $\cR_k$, which can also be understood from the behavior of $\vp_k$, since $\vp_{k,N}$ is still large enough in Fig. \ref{623}.

In summary, we compare our analytical approximation of the evolution of $\cR_k$ to the numerical calculation in our previous work in Ref. \cite{shuzhi}. Except the epoch around $N_{\rm out}$ with $k\sim He^N$, the theoretical and numerical results match each other very well. We admit that there is indeed some situations where our approximation is insufficient. These intricacies mainly root from the competition between the two terms in Eqs. (\ref{dayu}) and (\ref{dayun}), because it is not always easy to determine the leading-order one between them, so numerical calculation is still necessary there. Last, we claim that there are revolving processes of $\cR_k$ in all the five typical cases, but the subsequent linear evolutions only exist for the three intermediate scales, and the relevant criterion should be that $|\cR_k|$ changes dramatically and $\vp_{k,N}$ is tiny simultaneously.

\section{Conclusion} \label{sec:con}

The USR inflation is receiving increasing research interest in recent years, in which the primordial curvature perturbation $\cR_k$ behaves completely different in contrast to that in the SR case. The most remarkable character is that $\cR_k$ can still significantly increase after the horizon-exit and thus greatly enhance the power spectrum $\cP_\cR$ on small scales. As a result, PBHs at certain mass windows can be produced in abundance, serving as an effective candidate of DM. Meanwhile, SIGWs at certain frequencies can also be intense enough to be discovered by current and future gravitational wave detectors.

In this paper, we provide an analytical approximation to systematically investigate the evolution of ${\cal R}_k$ and the relevant $\cP_\cR$ in the USR inflation, which is usually a highly numerical and time-consuming task. The asymptotic solutions of the moduli and arguments of ${\cal R}_k$ and ${\cal R}_{k,N}$ are obtained in both sub- and super-horizon limits with $k\gg He^N$ and $k\ll He^N$, perfectly matching the numerical results in our previous work in Ref. \cite{shuzhi}. Moreover, five typical scales with different $N_{\rm out}$ are studied in order, corresponding to five different positions in $\cP_\cR$. In all, we hope to construct a framework to facilitate the analytical calculation of ${\cal R}_k$. Our basic conclusions are summarized as follows.

First, we obtain the approximate forms of the parameters $\ve$ and $\eta$ in both SR and USR stages. By this means, all the analytical solutions of $|{\cal R}_k|$, $|{\cal R}_k|_{,N}$, $\theta_{k,N}$, and $\vp_{k,N}$ can be expressed in simple exponential forms. However, due to the different asymptotic expansion of the Bessel function, the solutions with $k\ll He^N$ are more complicated than those with $k\gg He^N$, because we have to determine the leading-order term of $\cR_k$ in advance. All the possibilities are summarized in Tabs. \ref{table:1} and \ref{table:2}.

Second, as for the evolution of $\cR_k$ for the five typical scales with different $N_{\rm out}$, the numerical results validate our analytical approximation where it is applicable, as shown in Figs. \ref{fig:50}--\ref{fig:62}. The evolution of $\cR_k$ can basically be divided into two types: the revolving and linear processes. The former exists in all the five cases. At early times when $k\gg He^N$, in the first four cases with $N_{\rm out}<N_{\rm e}$, we have $|{\cal R}_k|\sim e^{-N}$ and $\theta_{k,N}\sim e^{-N}$, so ${\cal R}_k$ presents a revolving motion towards the origin in the complex plane with decreasing angular velocity, as shown in Figs. \ref{504}, \ref{52894}, \ref{554}, and \ref{58844}. However, in the last case with $N_{\rm out}>N_{\rm e}$, the USR stage causes $|{\cal R}_k|\sim e^{2N}$ and $\theta_{k,N}\sim e^{-N}$, so ${\cal R}_k$ revolves away from the origin, as shown in Fig. \ref{624}.

Third, for the three intermediate scales with $N_{\rm out}=52.89$, 55, and $58.84$, besides the revolving motion, a linear evolution of $\cR_k$ follows afterward. In these situations, $|\cR_k|$ changes violently, but at the same time $\vp_{k,N}$ is extremely small, so ${\cal R}_k$ nearly evolves along a straight line in the complex plane. When $N_{\rm out}=52.89$, ${\cal R}_k$ moves towards but not through the origin in Fig. \ref{52894}, inducing the sharp dip in $\cP_\cR$ but with a nonvanishing minimum. On the contrary, when $N_{\rm out}=55$ and $58.84$, $\cR_k$ departs away from the origin in Figs. \ref{554} and \ref{58844}, explaining the steep growth and the peak in $\cP_\cR$. However, if $N_{\rm out}$ is too small or too big, $\vp_{k,N}$ remains large enough during the USR stage, so there is only revolving process of $\cR_k$ without the subsequent linear evolution.

Altogether, by exploring the evolution of $\cR_k$ via our analytical approximation, we wish to provide a whole picture and thorough understanding of the primordial curvature perturbation and the power spectrum in the USR inflation from a theoretical perspective. Our work will be helpful to the model building of the USR inflation and contribute our knowledge of PBH and gravitational wave physics.

\acknowledgments

We are very grateful to G. Franciolini, B.-M. Gu, Y.-C. Liu, S. Pi, D. J. Schwarz, B.-Y. Su, and H.-R. Zhao for fruitful discussions. This work is supported by the Fundamental Research Funds for the Central Universities of China (No. N232410019).

\appendix

\section{Derivation of Eq. (\ref{th})} \label{appendix}

Below, we show the derivation of Eq. (\ref{th}) in more detail and prove that it is a general result.

For this purpose, it is more convenient to utilize the Mukhanov variable $v_k={\cal R}_kz$. By means of $v_k$, the MS equation in Eq. (\ref{MS}) can be rewritten as
\begin{flalign}
v_k''+\lt(k^2-\f{z''}{z}\rt)v_k=0, \label{MStau}
\end{flalign}
where a prime denotes the derivative with respect to the conformal time $\tau$ defined as
\begin{flalign}
\dd\tau=\f{\dd t}{a}=\f{\dd N}{He^N}. \label{gxsj}
\end{flalign}
Next, we decompose $v_k$ into its modulus and argument as
\begin{flalign}
v_k=|v_k|e^{-i\theta_k}. \label{vfenjie}
\end{flalign}
Since $z$ is real, $v_k$ has the same argument $\theta_k$ as that of ${\cal R}_k$.

From Eq. (\ref{vfenjie}), the first- and second-order derivatives of $v_k$ read
\begin{flalign}
v_k'&=(|v_k|'-i|v_k|\theta_k')e^{-i\theta_k}, \n\\ 
v_k''&=(|v_k|''-|v_k|\theta_k'^2-2i|v_k|'\theta_k'-i|v_k|\theta_k'')e^{-i\theta_k}. \n 
\end{flalign}
Substituting these results into Eq. (\ref{MStau}), it is straightforward to obtain $(|v_k|^2\theta_k')'=0$, 
indicating that $|v_k|^2\theta_k'$ is a constant. Equivalently, in terms of ${\cal R}_k$ and $N$, we have
\begin{flalign}
|{\cal R}_k|^2z^2\theta_{k,N}He^N={\rm const}. \label{Rc}
\end{flalign}
Here, we should emphasize that Eq. (\ref{Rc}) is general, as it is a direct result from the MS equation in essential. Consequently, it is valid in both SR and USR stages, and on both sub- and super-horizon scales.

Furthermore, we choose the Bunch--Davies vacuum as a special point to fix the constant in Eq. (\ref{Rc}). First, we rewrite ${\cal R}_k$ in Eq. (\ref{vk}) via the conformal time $\tau$ as
\begin{flalign}
{\cal R}_k=\f{e^{-ik\tau}}{z\sqrt{2k}}\lt(1-\f{i}{k\tau}\rt). \n
\end{flalign}
Therefore, we have 
\begin{flalign}
|{\cal R}_k|&=\sqrt{\f{1}{2z^2k}\lt[1+\f{1}{(k\tau)^2}\rt]}, \n\\ 
\theta_k&=k\tau+\arctan\f{1}{k\tau}. \n 
\end{flalign}
Substituting these results into Eq. (\ref{Rc}) and taking Eq. (\ref{gxsj}) into account, we can eventually determine the constant in Eq. (\ref{Rc}) as
\begin{flalign}
|{\cal R}_k|^2z^2\theta_{k,N}He^N=\f{1}{2}, \n
\end{flalign}
which is just Eq. (\ref{th}). 


\begin{thebibliography}{99}
\bibitem{LIGO}
B. P. Abbott {\it et al.} (LIGO Scientific and Virgo Collaboration), \emph{Observation of Gravitational
Waves from a Binary Black Hole Merger}, \emph{Phys. Rev. Lett.} {\bf 116} (2016) 061102, \arxgr{1602.03837}.

\bibitem{Bird:2016dcv}
S. Bird, I. Cholis, J. B. Mu\~noz, Y. Ali-Ha\"{\i}moud, M. Kamionkowski, E. D. Kovetz, A. Raccanelli, and A. G. Riess,
\emph{Did LIGO detect dark matter?}, \emph{Phys. Rev. Lett.} {\bf 116} (2016) 201301, \arxas{1603.00464}. 

\bibitem{jp}
M. Sasaki, T. Suyama, T. Tanaka, and S. Yokoyama, \emph{Primordial Black Hole Scenario for the Gravitational-Wave Event GW150914}, \emph{Phys. Rev. Lett.} {\bf 117} (2016) 061101, \arxas{1603.08338}.

\bibitem{Clesse}
S. Clesse and J. Garc\'{\i}a-Bellido, \emph{The clustering of massive Primordial Black Holes as Dark Matter: measuring their mass distribution with Advanced LIGO}, \emph{Phys. Dark Univ.} {\bf 15} (2017) 142, \arxas{1603.05234}. 


\bibitem{Green:2020jor}
A. M. Green and B. J. Kavanagh, \emph{Primordial Black Holes as a dark matter candidate}, \emph{J. Phys. G} {\bf 48} (2021) 043001, \arxas{2007.10722}.



\bibitem{Bowman:2018yin}
J. D. Bowman, A. E. E. Rogers, R. A. Monsalve, T. J. Mozdzen, and N. Mahesh,
\emph{An absorption profile centred at 78 megahertz in the sky-averaged spectrum}, \emph{Nature} \textbf{555} (2018) 67, \arxas{1810.05912}. 

\bibitem{Zhang:2023rnp}
Z. Zhang, B. Yue, Y. Xu, Y.-Z. Ma, X. Chen, and M. Liu, \emph{Cosmic Radio Background from Primordial Black Holes at Cosmic Dawn}, \emph{Phys. Rev. D} \textbf{107} (2023) 083013, \arxas{2303.06616}. 

\bibitem{pbh1}
S. Clesse and J. Garc\'{i}a-Bellido, \emph{Massive Primordial Black Holes from Hybrid Inflation as Dark Matter and the seeds of Galaxies}, \emph{Phys. Rev. D} {\bf 92} (2015) 023524, \arxas{1501.07565}.

\bibitem{2207.09436}
M. Castellano {\it et al.}, \emph{Early results from GLASS-JWST. III: Galaxy candidates at z$\sim$9-15}, \emph{Astrophys. J. Lett.} {\bf 938} (2022) L15, \arxas{2207.09436}.


\bibitem{Saito:2008jc}
R. Saito and J. Yokoyama, \emph{Gravitational wave background as a probe of the primordial black hole abundance}, \emph{Phys. Rev. Lett.} {\bf 102} (2009) 161101, \arxas{0812.4339}.

\bibitem{Domenech:2021ztg}
G. Dom\`enech, \emph{Scalar Induced Gravitational Waves Review}, \emph{Universe} {\bf 7} (2021) 398, \arxgr{2109.01398}. 

\bibitem{Ahmed:2021ucx}
W. Ahmed, M. Junaid, and U. Zubair, \emph{Primordial black holes and gravitational waves in hybrid inflation with chaotic potentials}, \emph{Nucl. Phys. B} \textbf{984} (2022) 115968, \arxas{2109.14838}. 

\bibitem{Kawai:2021edk}
S. Kawai and J. Kim, \emph{Primordial black holes from Gauss-Bonnet-corrected single field inflation}, \emph{Phys. Rev. D} \textbf{104} (2021) 083545, \arxas{2108.01340}. 

\bibitem{Lin:2021vwc}
J. Lin, S. Gao, Y. Gong, Y. Lu, Z. Wang, and F. Zhang,
\emph{Primordial black holes and scalar induced gravitational waves from Higgs inflation with noncanonical kinetic term}, \emph{Phys. Rev. D} \textbf{107} (2023) 043517, \arxgr{2111.01362}.

\bibitem{Yi:2020cut}
Z. Yi, Q. Gao, Y. Gong, and Z.-h. Zhu,
\emph{Primordial black holes and scalar-induced secondary gravitational waves from inflationary models with a noncanonical kinetic term}, \emph{Phys. Rev. D} \textbf{103} (2021) 063534, \arxas{2011.10606}.

\bibitem{Di:2017ndc}
H. Di and Y. Gong, \emph{Primordial black holes and second order gravitational waves from ultra-slow-roll inflation}, \emph{J. Cosmol. Astropart. Phys.} {\bf 07} (2018) 007, \arxas{1707.09578}.




\bibitem{NANOGrav:2023gor}
G. Agazie {\it et al.} (NANOGrav Collaboration), \emph{The NANOGrav 15 yr Data Set: Evidence for a Gravitational-wave Background}, \emph{Astrophys. J. Lett.} \textbf{951} (2023) L8, \arxas{2306.16213}. 

\bibitem{Antoniadis:2023ott}
J. Antoniadis {\it et al.}, \emph{The second data release from the European Pulsar Timing Array III. Search for gravitational wave signals}, \emph{Astron. Astrophys.} {\bf 678} (2023) A50, \arxas{2306.16214}.

\bibitem{Reardon:2023gzh}
D. J. Reardon {\it et al.}, \emph{Search for an Isotropic Gravitational-wave Background with the Parkes Pulsar Timing Array}, \emph{Astrophys. J. Lett.} \textbf{951} (2023) L6, \arxas{2306.16215}. 

\bibitem{Xu:2023wog}
H. Xu {\it et al.}, \emph{Searching for the Nano-Hertz Stochastic Gravitational Wave Background with the Chinese Pulsar Timing Array Data Release I}, \emph{Res. Astron. Astrophys.} \textbf{23} (2023) 075024, \arxas{2306.16216}. 

\bibitem{NANOGrav:2023hvm}
A. Afzal {\it et al.} (NANOGrav Collaboration), \emph{The NANOGrav 15 yr Data Set: Search for Signals from New Physics}, \emph{Astrophys. J. Lett.} \textbf{951} (2023) L11, \arxas{2306.16219}. 

\bibitem{Inomata:2023zup}
K. Inomata, K. Kohri, and T. Terada,
\emph{The Detected Stochastic Gravitational Waves and Subsolar-Mass Primordial Black Holes}, \emph{Phys. Rev. D} {\bf 109} (2024) 063506, \arxas{2306.17834}.

\bibitem{Wang:2023ost}
S. Wang, Z.-C. Zhao, J.-P. Li, and Q.-H. Zhu,
\emph{Implications of Pulsar Timing Array Data for Scalar-Induced Gravitational Waves and Primordial Black Holes: Primordial Non-Gaussianity $f_{\rm NL}$ Considered}, \arxas{2307.00572}.

\bibitem{Yi:2023mbm}
Z. Yi, Q. Gao, Y. Gong, Y. Wang, and F. Zhang, \emph{Scalar induced gravitational waves in light of Pulsar Timing Array data}, \emph{Sci. China Phys. Mech. Astron.} \textbf{66} (2023) 120404, \arxas{2307.02467}.

\bibitem{Firouzjahi:2023lzg}
H. Firouzjahi and A. Talebian, \emph{Induced Gravitational Waves from Ultra Slow-Roll Inflation and Pulsar Timing Arrays Observations}, \emph{J. Cosmol. Astropart.
Phys.} {\bf 10} (2023) 032, \arxas{2307.03164}.

\bibitem{Balaji:2023ehk}
S. Balaji, G. Dom\`enech, and G. Franciolini, \emph{Scalar-induced gravitational wave interpretation of PTA data: the role of scalar fluctuation propagation speed}, \emph{J. Cosmol. Astropart.
Phys.} {\bf 10} (2023) 041, \arxas{2307.08552}.

\bibitem{Franciolini:2023pbf}
G. Franciolini, A. J. Iovino, V. Vaskonen, and H. Veerm\"{a}e, \emph{The recent gravitational wave observation by pulsar timing arrays and primordial black holes: the importance of non-gaussianities}, \emph{Phys. Rev. Lett.} \textbf{131} (2023) 201401, \arxas{2306.17149}.

\bibitem{dm}
B. Carr and F. K\"{u}hnel, \emph{Primordial Black Holes as Dark Matter: Recent Developments}, \emph{Ann. Rev. Nucl. Part. Sci.} {\bf 70} (2020) 355, \arxas{2006.02838}.

\bibitem{Escriva:2022duf}
A. Escriv\`a, F. K\"{u}hnel, and Y. Tada, \emph{Primordial Black Holes}, \arxas{2211.05767}. 


\bibitem{Clesse:2017bsw}
S. Clesse and J.Garc\'{\i}a-Bellido, \emph{Seven Hints for Primordial Black Hole Dark Matter}, \emph{Phys. Dark Univ.} \textbf{22} (2018) 137, \arxas{1711.10458}. 

\bibitem{zs}
B. Carr, K. Kohri, Y. Sendouda, and J. Yokoyama, \emph{Constraints on primordial black holes}, \emph{Rept. Prog. Phys.} {\bf 84} (2021) 116902, \arxas{2002.12778}.

\bibitem{PS}
W. H. Press and P. Schechter, \emph{Formation of galaxies and clusters of galaxies by selfsimilar gravitational condensation}, \emph{Astrophys. J.} {\bf 187} (1974) 425.

\bibitem{peak}
J. M. Bardeen, J. R. Bond, N. Kaiser, and A. S. Szalay, \emph{The Statistics of Peaks of Gaussian Random Fields}, \emph{Astrophys. J.} {\bf 304} (1986) 15. 

\bibitem{Green:2004wb}
A. M. Green, A. R. Liddle, K. A. Malik, and M. Sasaki, \emph{A new calculation of the mass fraction of primordial black holes}, \emph{Phys. Rev. D} {\bf 70} (2004) 041502(R), \Arxas{0403181}.

\bibitem{Planck}
N. Aghanim {\it et al.} (Planck Collaboration), \emph{Planck 2018 results. VI. Cosmological parameters}, \emph{Astron. Astrophys.} {\bf 641} (2020) A6, \arxas{1807.06209}.

\bibitem{Garcia-Bellido:2017mdw}
J. Garc\'\i{}a-Bellido and E. R. Morales, \emph{Primordial black holes from single field models of inflation}, \emph{Phys. Dark Univ.} {\bf 18} (2017) 47, \arxas{1702.03901}.

\bibitem{Kannike:2017bxn}
K. Kannike, L. Marzola, M. Raidal, and H. Veerm\"ae, \emph{Single Field Double Inflation and Primordial Black Holes}, \emph{J. Cosmol. Astropart. Phys.} {\bf 09} (2017) 020, \arxas{1705.06225}. 

\bibitem{Germani:2017bcs}
C. Germani and T. Prokopec, \emph{On primordial black holes from an inflection point}, \emph{Phys. Dark Univ.} {\bf 18} (2017) 6, \arxas{1706.04226}. 

\bibitem{Motohashi:2017kbs}
H. Motohashi and W. Hu, \emph{Primordial Black Holes and Slow-Roll Violation}, \emph{Phys. Rev. D} {\bf 96} (2017) 063503, \arxas{1706.06784}.

\bibitem{Dimopoulos:2017ged}
K. Dimopoulos, \emph{Ultra slow-roll inflation demystified}, \emph{Phys. Lett. B} {\bf 775} (2017) 262, \arxph{1707.05644}.

\bibitem{Ezquiaga:2017fvi}
J. M. Ezquiaga, J. Garc\'\i{}a-Bellido, and E. R. Morales, \emph{Primordial Black Hole production in Critical Higgs Inflation}, \emph{Phys. Lett. B} {\bf 776} (2018) 345, \arxas{1705.04861}. 

\bibitem{Ballesteros:2017fsr}
G. Ballesteros and M. Taoso, \emph{Primordial black hole dark matter from single field inflation}, \emph{Phys. Rev. D} {\bf 97} (2018) 023501, \arxph{1709.05565}.

\bibitem{Cicoli:2018asa}
M. Cicoli, V. A. Diaz, and F. G. Pedro, \emph{Primordial Black Holes from String Inflation}, \emph{J. Cosmol. Astropart. Phys.} {\bf 06} (2018) 034, \arxth{1803.02837}. 

\bibitem{Ozsoy:2018flq}
O. \"Ozsoy, S. Parameswaran, G. Tasinato, and I. Zavala, \emph{Mechanisms for Primordial Black Hole Production in String Theory}, \emph{J. Cosmol. Astropart. Phys.} {\bf 07} (2018) 005, \arxth{1803.07626}. 

\bibitem{Dalianis:2018frf}
I. Dalianis, A. Kehagias, and G. Tringas, \emph{Primordial Black Holes from $\alpha$-attractors}, \emph{J. Cosmol. Astropart. Phys.} {\bf 01} (2019) 037, \arxas{1805.09483}. 

\bibitem{Ballesteros:2018wlw}
G. Ballesteros, J. B. Jim\'e{}nez, and M. Pieroni, \emph{Black hole formation from a general quadratic action for inflationary primordial fluctuations}, \emph{J. Cosmol. Astropart. Phys.} {\bf 06} (2019) 016, \arxas{1811.03065}. 

\bibitem{Cheng:2018qof}
S.-L. Cheng, W. Lee, and K.-W. Ng, \emph{Superhorizon curvature perturbation in ultra-slow-roll inflation}, \emph{Phys. Rev. D} {\bf 99} (2019) 063524, \arxas{1811.10108}.

\bibitem{Bhaumik:2019tvl}
N. Bhaumik and R. K. Jain, \emph{Primordial black holes dark matter from inflection point models of inflation and the effects of reheating}, \emph{J. Cosmol. Astropart. Phys.} {\bf 01} (2020) 037, \arxas{1907.04125}.

\bibitem{Mahbub:2019uhl}
R. Mahbub, \emph{Primordial black hole formation in inflationary $\alpha$-attractor models}, \emph{Phys. Rev. D} {\bf 101} (2020) 023533, \arxas{1910.10602}. 

\bibitem{liu}
J. Liu, Z.-K. Guo, and R.-G. Cai, \emph{An analytical approximation of the scalar spectrum in the ultra-slow-roll inflationary models}, \emph{Phys. Rev. D} {\bf 101} (2020) 083535, \arxas{2003.02075}.

\bibitem{Mishra:2019pzq}
S. S. Mishra and V. Sahni, \emph{Primordial Black Holes from a tiny bump/dip in the Inflaton potential}, \emph{J. Cosmol. Astropart. Phys.} {\bf 04} (2020) 007, \arxgr{1911.00057}.

\bibitem{Cai:2019bmk}
R.-G. Cai, Z.-K. Guo, J. Liu, L. Liu, and X.-Y. Yang, \emph{Primordial black holes and gravitational waves from parametric amplification of curvature perturbations}, \emph{J. Cosmol. Astropart. Phys.} {\bf 06} (2020) 013, \arxas{1912.10437}.

\bibitem{Figueroa:2020jkf}
D. G. Figueroa, S. Raatikainen, S. R\"as\"anen, and E. Tomberg, \emph{Non-Gaussian tail of the curvature perturbation in stochastic ultra-slow-roll inflation: implications for primordial black hole production}, \emph{Phys. Rev. Lett.} {\bf 127} (2021) 101302, \arxas{2012.06551}.


\bibitem{Ragavendra:2020sop}
H. V. Ragavendra, P. Saha, L. Sriramkumar, and J. Silk, \emph{PBHs and secondary GWs from ultra slow roll and punctuated inflation}, \emph{Phys. Rev. D} {\bf 103} (2021) 083510, \arxas{2008.12202}.

\bibitem{Choi:2021yxz}
K.-Y. Choi, S.-b. Kang, and R. N. Raveendran, \emph{Reconstruction of potentials of the hybrid inflation in the light of primordial black hole formation}, \emph{J. Cosmol. Astropart. Phys.} {\bf 06} (2021) 054, \arxas{2102.02461}.


\bibitem{Kefala:2020xsx}
K. Kefala, G. P. Kodaxis, I. D. Stamou, and N. Tetradis, \emph{Features of the inflaton potential and the power spectrum of cosmological perturbations}, \emph{Phys. Rev. D} {\bf 104} (2021) 023506, \arxas{2010.12483}.

\bibitem{Lyc}
Y.-C. Liu, Q. Wang, B.-Y. Su, and N. Li, \emph{Primordial black holes from the perturbations in the inflaton potential}, \emph{Phys. Dark Univ.} {\bf 34} (2021) 100905.


\bibitem{Wq}
Q. Wang, Y.-C. Liu, B.-Y. Su, and N. Li, \emph{Primordial black holes from the perturbations in the inflaton potential in peak theory}, \emph{Phys. Rev. D} {\bf 104} (2021) 083546, \arxas{2111.10028}.

\bibitem{Dalianis:2021iig}
I. Dalianis, G. P. Kodaxis, I. D. Stamou, N. Tetradis, and A. Tsigkas-Kouvelis, \emph{Spectrum oscillations from features in the potential of single-field inflation}, \emph{Phys. Rev. D} {\bf 104} (2021) 103510, \arxas{2106.02467}. 

\bibitem{Inomata:2021uqj}
K. Inomata, E. McDonough, and W. Hu, \emph{Primordial Black Holes Arise When The Inflaton Falls}, \emph{Phys. Rev. D} {\bf 104} (2021) 123553, \arxas{2104.03972}.

\bibitem{Cheng:2021lif}
S.-L. Cheng, D.-S. Lee, and K.-W. Ng, \emph{Power spectrum of primordial perturbations during ultra-slow-roll inflation with back reaction effects}, \emph{Phys. Lett. B} {\bf 827} (2022) 136956, \arxas{2106.09275}.

\bibitem{Wu:2021mwy}
Y.-P. Wu, E. Pinetti, K. Petraki, and J. Silk, \emph{Baryogenesis from ultra-slow-roll inflation}, \emph{J. High Energy Phys.} {\bf 01} (2022) 015, \arxas{2109.00118}.

\bibitem{Inomata:2021tpx}
K. Inomata, E. McDonough, and W. Hu, \emph{Amplification of primordial perturbations from the rise or fall of the inflaton}, \emph{J. Cosmol. Astropart. Phys.} {\bf 02} (2022) 031, \arxas{2110.14641}.

\bibitem{Figueroa:2021zah}
D. G. Figueroa, S. Raatikainen, S. R\"as\"anen, and E. Tomberg, \emph{Implications of stochastic effects for primordial black hole production in ultra-slow-roll inflation}, \emph{J. Cosmol. Astropart. Phys.} {\bf 05} (2022) 027, \arxas{2111.07437}.

\bibitem{Hooshangi:2022lao}
S. Hooshangi, A. Talebian, M. H. Namjoo, and H. Firouzjahi,
\emph{Multiple field ultraslow-roll inflation: Primordial black holes from straight bulk and distorted boundary}, \emph{Phys. Rev. D} {\bf 105} (2022) 083525, \arxas{2201.07258}.

\bibitem{Geller:2022nkr}
S. R. Geller, W. Qin, E. McDonough, and D. I. Kaiser, \emph{Primordial black holes from multifield inflation with nonminimal couplings}, \emph{Phys. Rev. D} {\bf 106} (2022) 063535, \arxth{2205.04471}.

\bibitem{Franciolini:2022pav}
G. Franciolini and A. Urbano, \emph{Primordial black hole dark matter from inflation: The reverse engineering approach}, \emph{Phys. Rev. D} {\bf 106} (2022) 123519, \arxas{2207.10056}.

\bibitem{Balaji:2022zur}
S. Balaji, H. V. Ragavendra, S. K. Sethi, J. Silk, and L. Sriramkumar, \emph{Observing Nulling of Primordial Correlations via the 21-cm Signal}, \emph{Phys. Rev. Lett.} {\bf 129} (2022) 261301, \arxas{2206.06386}.

\bibitem{Raveendran:2022dtb}
R. N. Raveendran, K. Parattu, and L. Sriramkumar,
\emph{Enhanced power on small scales and evolution of quantum state of perturbations in single and two field inflationary models},
\emph{Gen. Relativ. Gravit.} {\bf 54} (2022) 91, \arxas{2206.05760}.



\bibitem{Bhatt:2022mmn}
S. S. Bhatt, S. S. Mishra, S. Basak, and S. N. Sahoo, \emph{Numerical simulations of inflationary dynamics: slow-roll and beyond}, \arxgr{2212.00529}.

\bibitem{Gu:2022pbo}
B.-M. Gu, F.-W. Shu, K. Yang, and Y.-P. Zhang, \emph{Primordial black holes from an inflationary potential valley}, \emph{Phys. Rev. D} {\bf 107} (2023) 023519, \arxas{2207.09968}.

\bibitem{Zjx}
J.-X. Zhao, X.-H. Liu, N. Li, \emph{Primordial black holes and scalar-induced gravitational waves from the perturbations on the inflaton potential in peak theory}, \emph{Phys. Rev. D} {\bf 107} (2023) 043515, \arxas{2302.06886}. 

\bibitem{Mu:2022dku}
B. Mu, G. Cheng, J. Liu, and Z.-K. Guo, \emph{Constraints on ultraslow-roll inflation from the third LIGO-Virgo observing run}, \emph{Phys. Rev. D} {\bf 107} (2023) 043528, \arxas{2211.05386}.


\bibitem{Ragavendra:2023ret}
H. V. Ragavendra and L. Sriramkumar, \emph{Observational imprints of enhanced scalar power on small scales in ultra slow roll inflation and associated non-Gaussianities}, \emph{Galaxies} {\bf 11} (2023) 34, \arxas{2301.08887}.

\bibitem{Balaji:2022rsy}
S.~Balaji, J.~Silk, and Y.-P.~Wu, \emph{Induced gravitational waves from the cosmic coincidence}, \emph{J. Cosmol. Astropart. Phys.} \textbf{06} (2022) 008, \arxas{2202.00700}.

\bibitem{Qin:2023lgo}
W.~Qin, S.~R.~Geller, S.~Balaji, E.~McDonough, and D.~I.~Kaiser, \emph{Planck constraints and gravitational wave forecasts for primordial black hole dark matter seeded by multifield inflation}, \emph{Phys. Rev. D} \textbf{108} (2023) 043508, \arxas{2303.02168}.


\bibitem{Carrilho:2019oqg}
P. Carrilho, K. A. Malik, and D. J. Mulryne, \emph{Dissecting the growth of the power spectrum for primordial black holes}, \emph{Phys. Rev. D} {\bf 100} (2019) 103529, \arxas{1907.05237}.

\bibitem{Zhai:2023azx}
R. Zhai, H. Yu, and P. Wu, \emph{Power spectrum with $k^6$ growth for primordial black holes}, \emph{Phys. Rev. D} \textbf{108} (2023) 043529, \arxgr{2308.09286}. 

\bibitem{Franciolini:2023lgy}
G. Franciolini, A. Iovino, Jr., M. Taoso, and A. Urbano,
\emph{One loop to rule them all: Perturbativity in the presence of ultra slow-roll dynamics}, \arxas{2305.03491}.

\bibitem{Byrnes:2018txb}
C. T. Byrnes, P. S. Cole, and S. P. Patil, \emph{Steepest growth of the power spectrum and primordial black holes}, \emph{J. Cosmol. Astropart. Phys.} {\bf 06} (2019) 028, \arxas{1811.11158}. 

\bibitem{Ballesteros:2020sre}
G. Ballesteros, J. Rey, M. Taoso, and A. Urbano, \emph{Stochastic inflationary dynamics beyond slow-roll and consequences for primordial black hole formation}, \emph{J. Cosmol. Astropart. Phys.} {\bf 08} (2020) 043, \arxas{2006.14597}.

\bibitem{Tasinato:2020vdk}
G. Tasinato, \emph{An analytic approach to non-slow-roll inflation}, \emph{Phys. Rev. D} {\bf 103} (2021) 023535, \arxth{2012.02518}.

\bibitem{Ozsoy:2021qrg}
O. \"Ozsoy and G. Tasinato, \emph{CMB \ensuremath{\mu}T cross correlations as a probe of primordial black hole scenarios}, \emph{Phys. Rev. D} {\bf 104} (2021) 043526, \arxas{2104.12792}. 

\bibitem{Ozsoy:2021pws}
O. \"Ozsoy and G. Tasinato, \emph{Consistency conditions and primordial black holes in single field inflation}, \emph{Phys. Rev. D} {\bf 105} (2022) 023524, \arxas{2111.02432}.

\bibitem{Cole:2022xqc}
P. S. Cole, A. D. Gow, C. T. Byrnes, and S. P. Patil, \emph{Steepest growth re-examined: repercussions for primordial black hole formation}, \arxas{2204.07573}.

\bibitem{Karam:2022nym}
A. Karam, N. Koivunen, E. Tomberg, V. Vaskonen, and H. Veerm\"ae, \emph{Anatomy of single-field inflationary models for primordial black holes}, \emph{J. Cosmol. Astropart. Phys.} {\bf 03} (2023) 013, \arxas{2205.13540}. 

\bibitem{Pi:2022zxs}
S. Pi and J. Wang, \emph{Primordial Black Hole Formation in Starobinsky's Linear Potential Model}, \emph{J. Cosmol. Astropart. Phys.} {\bf 06} (2023) 018, \arxas{2209.14183}. 

\bibitem{shuzhi}
H.-R. Zhao, Y.-C. Liu, J.-X. Zhao, and N. Li,
\emph{The evolution of the primordial curvature perturbation in the ultraslow-roll inflation}, \emph{Eur. Phys. J. C} \textbf{83} (2023) 783.

\bibitem{Domenech:2023dxx}
G.~Dom\`enech, G.~Vargas, and T.~Vargas, \emph{An exact model for enhancing/suppressing primordial fluctuations}, \emph{J. Cosmol. Astropart. Phys.} \textbf{03} (2024) 002, \arxas{2309.05750}.

\bibitem{Alho:2020cdg}
A.~Alho, C.~Uggla, and J.~Wainwright, \emph{Dynamical systems in perturbative scalar field cosmology}, \emph{Classical Quantum Gravity} \textbf{37} (2020) 225011, \arxgr{2006.00800}.

\bibitem{KKLT}
S. Kachru, R. Kallosh, A. D. Linde, and S. P. Trivedi, \emph{de Sitter Vacua in String Theory}, \emph{Phys. Rev. D} {\bf 68} (2003) 046005, \Arxth{0301240}.

\bibitem{Mukhanov}
V. F. Mukhanov, \emph{, Quantum Theory of Gauge Invariant Cosmological Perturbations}, \emph{Sov. Phys. JETP} {\bf 68} (1988) 1297.

\bibitem{Sasaki}
M. Sasaki, \emph{Gauge Invariant Scalar Perturbations in the New Inflationary Universe}, \emph{Prog. Theor. Phys.} {\bf 76} (1986) 1036.

\bibitem{BD}
T. S. Bunch and P. C. W. Davies, \emph{Quantum Field Theory in de Sitter Space: Renormalization by Point Splitting}, \emph{Proc. Roy. Soc. Lond. A} {\bf 360} (1978) 117. 

\bibitem{Goswami:2010qu}
G. Goswami and T. Souradeep, \emph{Power spectrum nulls due to non-standard inflationary evolution}, \emph{Phys. Rev. D} \textbf{83} (2011) 023526, \arxas{1011.4914}. 
\end{thebibliography}
\end{document}